\documentclass{dcds-s}
\usepackage{amsmath}
\usepackage{paralist}
\usepackage{acronym}
\usepackage{graphics}
\usepackage{subfigure}
\usepackage{epsfig}
\usepackage[usenames,dvipsnames]{pstricks}
\usepackage{epsfig}
\usepackage{pst-grad} % For gradients
\usepackage{pst-plot} % For axes
\usepackage[colorlinks=true]{hyperref}
\hypersetup{urlcolor=blue, citecolor=red}

\def\currentvolume{3}
 \def\currentissue{2}
  \def\currentyear{2010}
   \def\currentmonth{June}
    \def\ppages{1--XX}
     \def\DOI{10.3934/dcdss.2010.3.xx}

\theoremstyle{definition}

\newtheorem{remark}{Remark}

\newcommand{\g}{\vec{g}}
\newcommand{\n}{\vec{n}}
\newcommand{\x}{\vec{x}}
\newcommand{\w}{\vec{w}}
\renewcommand{\t}{\vec{t}}
\renewcommand{\u}{\vec{u}}
\newcommand{\half}{{1\over2}}
\newcommand{\grad}{\nabla}
\renewcommand{\div}{\nabla\cdot}
\newcommand{\R}{\mathbb{R}}
\newcommand{\A}{\mathbb{A}}
\newcommand{\D}{\mathbf{D}}
\newcommand{\I}{\mathbf{I}}
\newcommand{\dt}{\partial_t}
\newcommand{\dx}{\partial_x}
\newcommand{\F}{\mathcal{F}}
\renewcommand{\H}{\mathcal{H}}
\newcommand{\Si}{\mathcal{S}}
\newcommand{\VOLNA}{\textsf{VOLNA }}
\newcommand{\OpenFOAM}{\textsf{OpenFOAM }}
\newcommand{\pd}[2]{\frac{\partial #1}{\partial #2}}

\acrodef{DNS}{Direct Numerical Simulation}
\acrodef{IVP}{Initial Value Problem}
\acrodef{NSWE}{Nonlinear Shallow Water Equations}
\acrodef{LSWE}{Linear Shallow Water Equations}

\title[On the relevance of \ac{NSWE}]%
      {On the relevance of the dam break problem in the context of nonlinear shallow water equations}

\author[Denys Dutykh and Dimitrios Mitsotakis]{}


 \email{Denys.Dutykh@univ-savoie.fr}
 \email{Dimitrios.Mitsotakis@math.u-psud.fr}

\thanks{The second author was supported by the Marie Curie Fellowship PIEF-GA-2008-219399 of the European Commission}

\begin{document}
\maketitle

\centerline{\scshape Denys Dutykh }
\medskip
{\footnotesize
 \centerline{Universit\'e de Savoie, CNRS-LAMA}
   \centerline{Campus Scientifique}
   \centerline{73376 Le Bourget-du-Lac France}
}

\medskip

\centerline{\scshape Dimitrios Mitsotakis }
\medskip
{\footnotesize
 \centerline{UMR de Math\'ematiques, Universit\'e de Paris-Sud}
   \centerline{B\^atiment 425, P.O. Box}
   \centerline{91405 Orsay France}
}

\bigskip

\begin{abstract}
The classical dam break problem has become the \textit{de facto} standard in validating the nonlinear shallow water equations solvers. Moreover, the \ac{NSWE} are widely used for flooding simulations. While applied mathematics community is essentially focused on developing new numerical schemes, we tried to examine the validity of the mathematical model under consideration. The main purpose of this study is to check the pertinence of the \ac{NSWE} for flooding processes. From the mathematical point of view, the answer is not obvious since all derivation procedures assumes the total water depth positivity. We performed a comparison between the two-fluid Navier-Stokes simulations and the \ac{NSWE} solved analytically and numerically. Several conclusions are drawn out and perspectives for future research are outlined.
\end{abstract}

%%%%%%%%%%%%%%%%%%%%%%%%%%%%%%%%%%%%%%%%%%%%%%%%%%%%%%%%%%%

\section{Introduction}

During the last century there were more than 200 failures of dams greater than 15 m high \cite{Singh1996, Zoppou2000}. They have caused a loss of more than 8000 lives and millions of dollars worth of damage. Consequently, dam break flows have become an important practical problem in civil engineering. Numerical models have become essential as a predictive tool in evaluating the risks associated with the failure of the hydraulic structures. That is why, the number of numerical studies has drastically increased during past decades.

To our knowledge, the dam break problem was studied analytically for the first time in the PhD thesis of Pohle (1950), \cite{Pohle1950}, who used a lagrangian description to solve this problem. The classical analytical solution for the dam break problem in the context of  the \ac{NSWE} can be found in the book of Stoker (1957), \cite{Stoker1957}. Later, this solution was generalized to the constant slope case by Mangeney {\sl et al}. (2000), \cite{Mangeney2000}. Note, that Hunt (1982), \cite{Hunt1982}, also considered the sloping channel case and he obtained a closed-form solution using a kinematic wave approximation. Among classical works on this topic, we have to mention the prominent paper by Benjamin (1968), \cite{Benjamin1968}. Recently, Korobkin \& Yilmaz (2008), \cite{Korobkin2008}, studied the initial stages of the dam break flow in the framework of potential free surface flows.

It is interesting, however, to recall some other known analytical solutions to \ac{NSWE} even if they are not directly related to the dam break problem. Wave run-up on a sloping beach was investigated by Carrier \& Greenspan (1958), \cite{CG58, CWY}, using a hodograph transformation\footnote{In this case, the hodograph transformation means that Riemann invariants were chosen as independent variables. After this change of variables, governing equations become linear and they are further solved by Hankel transform.}. This solution is extensively used in the tsunami waves community to validate the run-up algorithm of various \ac{NSWE} solvers \cite{Imamura1996, Titov1997, TS, noaa_report, Syno2006, Imamura2006, Dutykh2009a}. The transform of Carrier \& Greenspan was employed later by Synolakis and his collaborators to study analytically tsunami run up on a sloping beach, cf. e.g. \cite{Synolakis1987, Tadepalli1994, Tadepalli1996, Kanoglu2006}. There is also an analytical solution by Liu {\sl et al}. (2003), \cite{Liu2003} of the linearized shallow water equations on a sloping beach, where they used a forcing term to model an underwater landslide. This solution is also currently used to test numerical codes, \cite{Dutykh2009a}.

On the other hand, the dam break problem and various lock exchange flows were extensively studied experimentally, cf. e.g. \cite{Martin1952, Wood1970, Keller1991, Hacker1996, Stansby1998, Blaise2000, Bellos2004, Shin2004}. In this study, we do not directly appeal to them, since our main concern is to study the validity of \ac{NSWE} as an approximation to more complex mathematical models in some extreme situations.

Numerical studies are also countless. We can divide them conventionally into two big groups. In the first group, authors solved this problem in the framework of the \ac{NSWE}, cf. e.g. \cite{Wubs1988, Anastasiou1999, Tseng1999, Zoppou2000, Hervouet2000, Causon2000, Zhou2002, Gottardi2004, Bookhove2005, Xing2005, Benkhaldoun2006, Bai2007, Poncet2008} and in the second one where more advanced models were used, cf. e.g. \cite{Harlow1965, Nichols1971, Hirt1981, Hartel2000, Shao2003, Ozgokmen2004, Birman2005, Ooi2006, Ozgokmen2006, Bongolan-Walsh2007, Ozgokmen2007, Chang2008}. Obviously, this list does not pretend to be exhaustive.

The authors decided to perform this study because there is an apparent contradiction between the mathematical origins of the \ac{NSWE} and some applications of this model. When we look carefully at any derivation procedure of \ac{NSWE}, we will see that an implicit assumption of water depth positivity is adopted. Moreover, these equations are designed to model infinitely long waves. That is why, strictly speaking, these equations can be valid only in fluid regions. However, using various numerical techniques (sometimes ad-hoc, semiempirical) this model is routinely used for wetting/drying (run-up/run-down) simulations, cf. e.g. \cite{Titov1997, TS}. This process is considerably more complex and the validity of the \ac{NSWE} is not obvious \textit{a priori}. Recall, that the shoreline can be considered as a triple point: water, air and solid (soil) meet their. Of course, this situation is simplified for mathematical modelling.

We choose a \ac{DNS} by the two-fluid Navier-Stokes equations \cite{Scardovelli1999, Popinet1999} as the reference solution. This system contains all the necessary physical effects ranging from viscosity to the surface tension. Moreover, the ambient fluid (air) is resolved. In the absence of experimental data, these simulations can be assimilated to an idealized experiment. Up to graphical resolution, our numerical results are very similar to the experiments of J. Martin and W. Moyce \cite{Martin1952} and we remain clearly in the laminar r\'egime. We also underline that we consider a realistic density ratio 1:1000 as for the air/water interface (see Table \ref{tab:params}). The results of the \ac{DNS} are compared with several solutions to the \ac{NSWE}. Namely, the analytical solution of Stoker \cite{Stoker1957} (see Section \ref{sec:stoker}) was used in our comparison. Numerical solutions to the \ac{NSWE} were obtained using the \VOLNA code, cf. \cite{Dutykh2007a, Poncet2008, Dutykh2009a}.

The present study is organized as follows. In Section \ref{sec:models} we present two mathematical models which are used in this study. In the same section we also discuss several mathematical properties and extensions of the \ac{NSWE}. In Section \ref{sec:analytics} we review some known analytical solutions to the \ac{NSWE} of the dam break problem. After discussing briefly the numerical techniques, (Section \ref{sec:num}), we present and discuss our numerical results in Section \ref{sec:compare}. Conclusions are outlined in Section \ref{sec:concl}.

%%%%%%%%%%%%%%%%%%%%%%%%%%%%%%%%%%%%%%%%%%%%%%%%%%%%%%%%%%%

\section{Mathematical models}\label{sec:models}

In this section we briefly present two mathematical models which are used in the sequel. The first model is the well-known \acf{NSWE} which were derived for the first time by Saint-Venant (1871), cf. \cite{SV1871}. The second model is the two-fluid Navier-Stokes equations written under the assumption of fluids immiscibility. These equations are much more complete from physical and mathematical points of view. That is why, the two-fluid model is supposed to provide us reliable results.

\subsection{\acl{NSWE}}

The \acl{NSWE} can be written in the following conservative form (2DH):
\begin{align}
  H_t + \div (H\u) &= 0, \label{eq:gov1} \\
  (H\u)_t + \div\Bigl(H\u\otimes\u + \frac{g}{2}H^2\I\Bigr) &= gH\grad h, \label{eq:gov2}
\end{align}
where $H(\x,t)$ is the total water depth and $\u (\x,t):\R^2\times\R^+\mapsto \R^2$ is the depth-averaged horizontal velocity. Traditionally, $g$ denotes the acceleration due to the gravity, $h(\x,t)$ is the bathymetry function and $\I$ is the identity tensor.

We do not provide here the derivation of these equations since it is more than classical and can be found in various sources \cite{Stoker1957, Mei1994}.

\begin{remark}
The bathymetry function $h (\x,t)$ can be time-dependent. It is especially important for tsunami generation problems by submarine earthquakes, landslides, etc. The coupling with seismology is usually done through this function. Namely, various earthquake models, cf. e.g. \cite{Dutykh2006, Dutykh2007b, Dutykh2008, Kervella2007} give us the seabed displacements which are then transmitted to the ocean layer. Obviously, in this study we consider the fluid propagation over the flat bottom in view of applying analytical techniques.
\end{remark}

Governing equations (\ref{eq:gov1}), (\ref{eq:gov2}) form the system of balance laws (conservation laws, if the bottom is even $h = \textrm{const}$). Moreover, this system is strictly hyperbolic provided that $H > 0$. This property is extensively used in the construction of various numerical schemes and, in particular, in the Characteristic Flux approach, cf. \cite{Ghidaglia1996, Ghidaglia1998, Ghidaglia2001a, Ghidaglia2001, Dutykh2009a}, which is also implemented in the code \VOLNA.

Let us discuss the eigensystem of the advective flux. First, we introduce the so-called conservative variables and rewrite the governing equations as a system of conservation laws:
\begin{equation}\label{eq:diffsys}
  \pd{\w}{t} + \div\F(\w) = \Si (\w),
\end{equation}
where we introduced the following notations:
\begin{equation*}
  \w(\x,t):\R^2\times\R^+\mapsto \R^3, \qquad \w = (w_1, w_2, w_3) = (H, Hu, Hv),
\end{equation*}
\begin{equation*}
  \F (\w) = \begin{pmatrix}
    Hu & Hv \\
    Hu^2 + \frac{g}{2}H^2 & Huv \\
    Huv & Hv^2 + \frac{g}{2}H^2 \\
  \end{pmatrix} =
  \begin{pmatrix}
    w_2 & w_3 \\
    \frac{w_2^2}{w_1} + \frac{g}{2}w_1^2 & \frac{w_2w_3}{w_1} \\
    \frac{w_2w_3}{w_1} & \frac{w_3^2}{w_1} + \frac{g}{2}w_1^2 \\
  \end{pmatrix}.
\end{equation*}

After projecting the flux $\F(\w)$ on a unit normal direction $\n = (n_x, n_y)$, $|\n|=1$, one can compute the Jacobian matrix $\A_n$. Its expression in the physical variables has the following form:
\begin{equation*}
  \A_n = \pd{\bigl(\F(\w)\cdot\n\bigr)}{\w} =
  \begin{pmatrix}
    0 & n_x & n_y \\
    -u u_n + gHn_x & u_n + u n_x & u n_y \\
    -v u_n + gHn_y & v n_x & u_n + v n_y \\
  \end{pmatrix},
\end{equation*}
where $u_n = \u\cdot\n$ is the velocity vector projected on $\n$. The Jacobian matrix $\A_n$ has three distinct eigenvalues:
\begin{equation}\label{eq:eigen}
  \lambda_1 = u_n - c, \quad \lambda_2 = u_n, \quad \lambda_3 = u_n + c,
\end{equation}
where $c = \sqrt{gH}$ is the gravity wave speed in infinite wavelength limit. This quantity plays the same r\^ole as the speed of sound in compressible fluid mechanics. The hyperbolicity condition for the system (\ref{eq:gov1}), (\ref{eq:gov2}) follows immediately from (\ref{eq:eigen}) and the definition of $c$. The eigenstructure of the Jacobian matrix $\A_n$ is fundamental for constructing numerical flux function, cf. \cite{Dutykh2009a}, and thus, upwinding the discrete solution.

\subsubsection{Properties}

\acl{NSWE} have many other interesting properties. Some of them will be briefly recalled here. To reveal these properties, we shall take the water wave theory point of view.

Let us recast equations (\ref{eq:gov1}), (\ref{eq:gov2}) in the following nonconservative form in one space dimension:
\begin{align}\label{eq:eta}
  \dt\eta + \dx\bigl((h+\eta)u\bigr) &= 0, \\
  \dt u + \frac12\dx|u|^2 + g\dx\eta &= 0. \label{eq:u}
\end{align}
These equations possess a (non-canonical) Hamiltonian structure \cite{Salmon1988, Zakharov1997, Radder1999}:
\begin{equation*}
  \dt\begin{pmatrix}
    \eta \\
    u
  \end{pmatrix} +
  \begin{pmatrix}
    0 & \dx \\
    \dx & 0
  \end{pmatrix}
  \begin{pmatrix}
    \frac{\delta\H}{\delta\eta} \\
    \frac{\delta\H}{\delta u}
  \end{pmatrix} = 0,
\end{equation*}
where the Hamiltonian $\H$ is defined as
\begin{equation*}
  \H := \frac12\int\limits_{-\infty}^{+\infty} g\eta^2\;dx +
  \frac12\int\limits_{-\infty}^{+\infty} (h+\eta) u^2\;dx.
\end{equation*}
Moreover, the pair of equations (\ref{eq:eta}), (\ref{eq:u}) possesses an infinity of conservation laws \cite{Benney1974, Miura1974}.

Equations (\ref{eq:eta}), (\ref{eq:u}) can be also derived from Luke's Lagrangian variational principle \cite{Luke1967} if we introduce the velocity potential function $\phi (\x,t)$ such that $\u = \grad\phi$. In this case, the Lagrangian reads
\begin{equation*}
  \mathcal{L} = \int\limits_{t_1}^{t_2}\int\limits_{\x_1}^{\x_2} \Bigl\{(\eta + h)\bigl(\phi_t + \frac12|\grad\phi|^2\bigr) + \frac12 g\eta^2\Bigr\} \;d\x\;dt.
\end{equation*}
Governing equations (\ref{eq:eta}), (\ref{eq:u}) are obtained by varying $\mathcal{L}$ with respect to $\eta$ and $\phi$. Recently, a generalized variational principle was proposed by Clamond \& Dutykh, cf. \cite{Clamond2009}. Their approach allows for more flexibility and can be used to derive various generalized shallow and deep water approximations.

\subsubsection{Extensions}

The \acl{NSWE} (\ref{eq:gov1}), (\ref{eq:gov2}) arise after a series of approximations applied to complete set of equations. Strictly speaking, they model the propagation and transformation of infinitely long water waves. That is why, their validity for run-up and flooding simulations is not so obvious \textit{a priori}.

The validity region of these equations can be extended by adding some new physical effects. The inclusion of the dispersion is beneficial for description of shorter wavelengths. As a result, one can derive Boussinesq equations, \cite{Boussinesq1871, Boussinesq1872, Madsen03, BCS, Bona2004, BCL, Bona2007}, Serre equations, \cite{Serre1953, Barthelemy2004}, Green-Naghdi model, \cite{Green1974, Green1976, Kim2001, Li2002}, and several others.

Another physical effect is the dissipation. Situations where dissipation becomes important for water waves are discussed in \cite{Zabusky1971, Bona1981, Wu1981, DutykhDias2007, Dutykh2008a, Dutykh2008b}. If one neglects the bottom boundary layer effects \cite{DutykhDias2007, Dutykh2008a}, dissipative equations (\ref{eq:eta}), (\ref{eq:u}) take the following form, cf. \cite{Dias2007, DutykhDias2007, Dutykh2008a}:
\begin{align*}
  \dt\eta + \div\bigl((h+\eta)\u\bigr) &= \nu\grad^2\eta, \\
  \dt\u + \frac12\grad|\u|^2 + g\grad\eta &= \nu\grad^2\u,
\end{align*}
where $\nu$ is the kinematic viscosity. Corresponding dissipative Boussinesq equations can be found in \cite{Khabakhpashev1997, Liu2004, Liu2006, DutykhDias2007, Dutykh2008a}.

When equations are recast in the conservative form (\ref{eq:gov1}), (\ref{eq:gov2}), there is also an alternative approach to include the dissipation initiated by Gerbeau \& Perthame and followed by other authors, cf. e.g. \cite{Gerbeau2001, Bresch2003, Bresch2006, Marche2007}.

\subsection{Two-fluid Navier-Stokes equations}

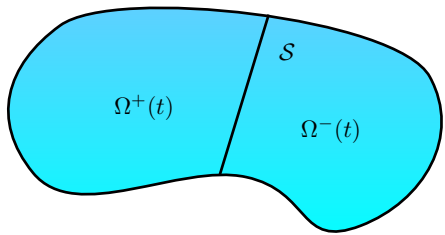
\begin{figure}
    \scalebox{0.9} % Change this value to rescale the drawing.
    {
    \begin{pspicture}(0,-1.9729664)(6.9945483,1.9723125)
    \definecolor{color1g}{rgb}{0.4,0.8,1.0}
    \psbezier[linewidth=0.04,fillstyle=gradient,gradlines=2000,gradbegin=color1g,gradmidpoint=1.0](3.7758446,-0.87319964)(4.7539034,-1.0815277)(4.5428023,-1.9529666)(5.4758444,-1.5931996)(6.408887,-1.2334328)(6.9745483,-0.27117625)(6.4958444,0.6068004)(6.017141,1.484777)(1.8465083,1.9523125)(1.0358446,1.3668003)(0.22518091,0.7812883)(0.0030853755,-0.038851038)(0.6358446,-0.81319964)(1.2686038,-1.5875483)(2.7977855,-0.66487163)(3.7758446,-0.87319964)
    \psline[linewidth=0.04cm](3.3958447,-0.8331996)(4.1158447,1.5268004)
    \usefont{T1}{ptm}{m}{n}
    \rput(2.2772508,0.15680036){$\Omega^+ (t)$}
    \usefont{T1}{ptm}{m}{n}
    \rput(5.037251,-0.18319964){$\Omega^- (t)$}
    \usefont{T1}{ptm}{m}{n}
    \rput(4.387251,0.9768004){$\Si$}
    \end{pspicture}
    }
    \caption{Two immiscible fuids separated by an interface.}
    \label{fig:omega}
\end{figure}

Let us consider two immiscible and incompressible\footnote{The case of the two compressible and miscible fluids was recently studied by Dias, Dutykh and Ghidaglia, cf. \cite{Dutykh2007a, Dias2008, Dias2008a, Dias2008b}.} fluids (water and air, for example) occupying domain $\Omega = \Omega^+ \cup \Omega^-$, where they are separated by an interface $\Si$. This situation is schematically depicted in Figure~\ref{fig:omega}. We note that we do not make any assumption on the interface complexity and topology. In what follows we will denote by superscripts $\pm$ all quantities related to the heavy and light fluids respectively.

In each fluid we can write mass and momentum balance equations:
\begin{eqnarray}
  \div\u &=& 0, \label{eq:mass_pm} \\
  \rho^\pm(\dt\u + \u\cdot\grad\u) + \grad p &=& \div(2\mu^\pm\D)
  + \sigma\kappa\delta_\Si\n + \rho^\pm\g. \label{eq:moment_pm}
\end{eqnarray}
The latter may be written in conservative form:
\begin{equation*}
  \dt(\rho^\pm\u) + \div(\rho^\pm\u\otimes\u + p\I) =
  \div(2\mu^\pm\D) + \sigma\kappa\delta_\Si\n + \rho^\pm\g,
\end{equation*}
where $\u$ is the fluid velocity, $\rho^\pm$ are the fluids densities, $\mu^\pm$ are the fluids dynamic viscosities, $\D = \half(\partial_i u_j + \partial_j u_i)$ is the rate of deformation tensor. The surface tension term is a force concentrated at the interface, $\sigma$ is the surface tension coefficient, $\kappa$ is the curvature of the interface, $\n$ is the unit normal to the interface and $\delta_\Si$ is the distribution (Dirac mass function) concentrated on the interface $\Si$.

Governing equations (\ref{eq:mass_pm}), (\ref{eq:moment_pm}) have to be completed by the following jump conditions across the interface:
\begin{itemize}
  \item Velocity continuity
  \begin{equation}\label{eq:velcont}
    [\u]_\Si = 0
  \end{equation}

  \item Tangential stress condition
  \begin{equation}
    [\mu\t\cdot\D\cdot\n]_\Si = 0,
  \end{equation}

  \item Normal stress condition
  \begin{equation}\label{eq:stress}
    [\n\cdot(-p\I + 2\mu\D)\cdot\n]_\Si = \sigma\kappa,
  \end{equation}
\end{itemize}
where $\t$ is a tangent vector ($\t\cdot\n = 0$) to the interface and notation $[\cdot]_\Si$ represents the jump of a quantity across the surface $\Si$.

However, for numerical computations it is advantageous to introduce a characteristic function $\phi$, (cf. \cite{Ishii1975, Toumi1996, Ghidaglia2001, Dias2008, Dias2008a, Dias2008b}) defined as:
\begin{equation*}
  \phi = \left\{
    \begin{array}{ll}
      1, & \x \in \Omega^+ (t), \\
      0, & \x \in \Omega^- (t). \\
    \end{array}
  \right.
\end{equation*}
Thus, $\phi$ and $\n$ are related by the formula $\grad\phi = \n\delta_\Si$. In the absence of phase change, $\phi$ is simply advected by the fluid motion:
\begin{equation}\label{eq:advect}
  \dt\phi + \div(\phi\u) = 0.
\end{equation}
In order to write a unique formulation for the entire domain, we express the density and the viscosity as functions of $\phi$:
\begin{equation*}
  \rho = \phi\rho^+ + (1-\phi)\rho^-, \quad
  \mu = \phi\mu^+ + (1-\phi)\mu^-.
\end{equation*}
Thus, we have the following momentum balance equation:
\begin{equation}\label{eq:momentum}
  \rho(\dt\u + \u\cdot\grad\u) + \grad p = \div(2\mu\D)
  + \sigma\kappa\delta_\Si\n + \rho\g.
\end{equation}
Along with the mass conservation equation (\ref{eq:mass_pm}) and the volume fraction advection equation (\ref{eq:advect}), it forms the two-fluid Navier-Stokes equations with an interface, which are solved numerically below.

\begin{remark}
We can recover jump conditions (\ref{eq:velcont}) -- (\ref{eq:stress}) if we investigate the governing equations (\ref{eq:mass_pm}), (\ref{eq:advect}), (\ref{eq:momentum}) in the neighborhood of the surface $\Si$ and making use of the formula $\grad\phi = \n\delta_\Si$.
\end{remark}

%%%%%%%%%%%%%%%%%%%%%%%%%%%%%%%%%%%%%%%%%%%%%%%%%%%%%%%%%%%

\section{Analytical solutions}\label{sec:analytics}

In this section we review known analytical solutions related to the dam break problem that we use in the comparison with the numerical results.

\subsection{Linear solution}

\begin{figure}
    \scalebox{0.9} % Change this value to rescale the drawing.
    {
    \begin{pspicture}(0,-1.7029687)(11.821875,1.7029687)
    \definecolor{color125b}{rgb}{0.8,0.8,1.0}
    \psframe[linewidth=0.0020,linecolor=color125b,dimen=outer,fillstyle=solid,fillcolor=color125b](6.84,-0.5754688)(4.4,-1.1754688)
    \psframe[linewidth=0.0040,linecolor=color125b,dimen=outer,fillstyle=solid,fillcolor=color125b](11.42,0.00453125)(6.82,-1.1754688)
    \psline[linewidth=0.02cm,arrowsize=0.05291667cm 2.0,arrowlength=1.4,arrowinset=0.4]{->}(0.0,-1.1554687)(11.64,-1.1554687)
    \psline[linewidth=0.02cm,arrowsize=0.05291667cm 2.0,arrowlength=1.4,arrowinset=0.4]{->}(5.64,-1.1554687)(5.64,1.6445312)
    \psline[linewidth=0.04cm](5.66,-0.55546874)(6.82,-0.55546874)
    \psline[linewidth=0.04cm](6.82,-0.55546874)(6.8,0.02453125)
    \psline[linewidth=0.04cm](6.8,0.04453125)(11.4,0.04453125)
    \psline[linewidth=0.04cm](5.64,-0.55546874)(4.4,-0.55546874)
    \psline[linewidth=0.04cm](4.4,-0.5754688)(4.4,-1.1554687)
    \psline[linewidth=0.04cm,arrowsize=0.05291667cm 2.0,arrowlength=1.4,arrowinset=0.4]{->}(6.82,-0.27546874)(7.64,-0.27546874)
    \psline[linewidth=0.04cm,arrowsize=0.05291667cm 2.0,arrowlength=1.4,arrowinset=0.4]{->}(4.38,-0.87546873)(3.62,-0.87546873)
    \usefont{T1}{ptm}{m}{n}
    \rput(11.521406,-1.5254687){$x$}
    \usefont{T1}{ptm}{m}{n}
    \rput(5.291406,1.5145313){$y$}
    \usefont{T1}{ptm}{m}{n}
    \rput(3.3014061,-0.84546876){$-c_0$}
    \usefont{T1}{ptm}{m}{n}
    \rput(7.9014063,-0.30546874){$c_0$}
    \psline[linewidth=0.03cm,arrowsize=0.05291667cm 2.0,arrowlength=1.4,arrowinset=0.4]{<->}(10.16,-0.03546875)(10.16,-1.1354687)
    \usefont{T1}{ptm}{m}{n}
    \rput(10.491406,-0.62546873){$h_0$}
    \usefont{T1}{ptm}{m}{n}
    \rput(5.5914063,-1.5054687){$O$}
    \psline[linewidth=0.03cm,arrowsize=0.05291667cm 2.0,arrowlength=1.4,arrowinset=0.4]{<->}(6.28,-0.59546876)(6.28,-1.1354687)
    \usefont{T1}{ptm}{m}{n}
    \rput(6.771406,-0.8854687){$h_0/2$}
    \end{pspicture}
    }
    \caption{Sketch of the solution to the \ac{LSWE}.}
    \label{fig:lswe}
\end{figure}
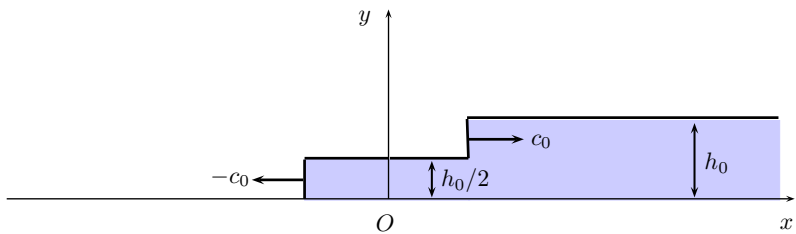

The simplest analytical solution for the dam break problem can be derived when we consider \ac{LSWE}. The latter can be obtained in a straightforward manner from (\ref{eq:eta}), (\ref{eq:u}):
\begin{align*}
  \dt\eta + \dx(h_0 u) &= 0, \\
  \dt u + g\dx\eta &= 0.
\end{align*}
In some situations, it is advantageous to eliminate the velocity variable $u$ to obtain
\begin{equation}\label{eq:etaWave}
  \pd{^2\eta}{t^2} - \pd{}{x}\Bigl(c_0^2\pd{\eta}{x}\Bigr) = 0, \quad
  c_0 := \sqrt{gh_0}.
\end{equation}
The \ac{IVP} for (\ref{eq:etaWave}) corresponding to the dam break takes the following form:
\begin{equation*}
  \eta (x, 0) = h_0\H(x), \quad \dt\eta (x,0) = 0,
\end{equation*}
where $\H(x)$ is the Heaviside function. This \ac{IVP} can be easily solved using the Fourier transform:
\begin{equation*}
  \eta (x,t) = h_0\Bigl(\frac12 +
  \frac{1}{\pi}\int\limits_0^{+\infty}\frac{\sin(kx)}{k}\cos(c_0kt)\;dk\Bigr).
\end{equation*}
The sketch of this solution is presented in Figure \ref{fig:lswe}. Namely, it consists of two waves propagating in opposite directions with velocities $\pm c_0$. Hence, the front speed is equal to $-c_0$. Of course, this result is nonphysical as it will be shown below.

Similar solutions can be constructed considering the linearized Euler equations for either one or two fluids separated by an interface.

\subsection{Small time asymptotics}

Several small time asymptotics were proposed to solve the dam break problem. One of the first solutions was derived by Pohle (1950), \cite{Pohle1950}. Such methods generally require the use of lagrangian description. The prominent book by Stoker, \cite{Stoker1957}, also contains such a solution:
\begin{align}
  X (a,b,t) &= a - \frac{g}{2\pi}t^2
  \log\left(\frac{\cos^2\frac{\pi b}{4h_0} + \sinh^2\frac{\pi a}{4h_0}}
  {\sin^2\frac{\pi b}{4h_0} + \sinh^2\frac{\pi a}{4h_0}}\right)+o(t^2),\label{eq:X} \\
  Y (a,b,t) &= b - \frac{g}{\pi}t^2
  \arctan\left(\frac{\sin\frac{\pi b}{2h_0}}{\sinh\frac{\pi a}{2h_0}}\right) + o(t^2), \label{eq:Y}
\end{align}
where $(X,Y)$ are new coordinates of the particle $(a,b)$ at time $t$. Recently this solution was generalized by Korobkin \& Oguz (2008) \cite{Korobkin2008}. We tried to compare the solution (\ref{eq:X}), (\ref{eq:Y}) with our numerical results and found that its validity time is too short for any practical use. That is why this solution is not plotted bellow.

Note, that expressions (\ref{eq:X}) and (\ref{eq:Y}) are singular at the shoreline $(a,b) = (0,0)$. Thus, some special care is needed to get an asymptotic expansion valid in the vicinity of this point, cf. \cite{Korobkin2008}.

\subsection{Nonlinear solution}\label{sec:stoker}

\begin{figure}
    \scalebox{0.9} % Change this value to rescale the drawing.
    {
    \begin{pspicture}(0,-1.82375)(11.441875,1.80675)
    \definecolor{color89g}{rgb}{0.8,0.8,1.0}
    \definecolor{color89f}{rgb}{0.4,0.8,1.0}
    \psframe[linewidth=0.0020, linecolor=white, dimen=outer, fillstyle=gradient, gradlines=2000, gradbegin=color89g, gradend=color89f, gradmidpoint=1.0] (10.72,0.46375)(5.42,-1.21625)
    \usefont{T1}{ptm}{m}{n}
    \rput(2.4114063,-0.50625){$H\equiv 0, u\equiv 0$}
    \psline[linewidth=0.046cm,arrowsize=0.05291667cm 2.0,arrowlength=1.4,arrowinset=0.4]{->}(0.0,-1.23625)(11.36,-1.23625)
    \usefont{T1}{ptm}{m}{n}
    \rput(11.141406,-1.64625){$x$}
    \psline[linewidth=0.046cm,arrowsize=0.05291667cm 2.0,arrowlength=1.4,arrowinset=0.4]{<-}(5.42,1.78375)(5.42,-1.23625)
    \usefont{T1}{ptm}{m}{n}
    \rput(7.731406,-0.40625){$H\equiv h_0, u\equiv 0$}
    \usefont{T1}{ptm}{m}{n}
    \rput(5.351406,-1.56625){$O$}
    \usefont{T1}{ptm}{m}{n}
    \rput(4.9514065,1.57375){$y$}
    \psline[linewidth=0.046cm](5.42,0.48375)(10.72,0.48375)
    \end{pspicture}
    }
    \caption{Sketch of the initial condition for the shallow water computations.}
    \label{fig:initcond}
\end{figure}
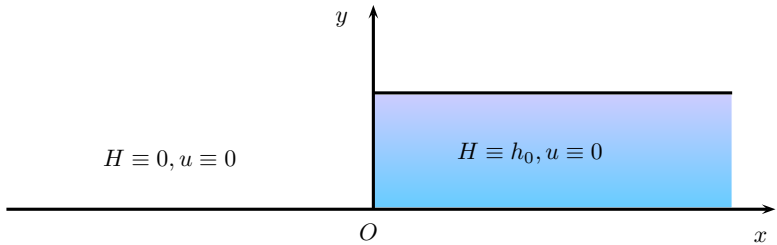

The classical book by J.J. Stoker, \cite{Stoker1957}, contains an analytical solution of the NSWE for the dam break problem. Consider the classical initial condition:
\begin{equation*}
  H (x, 0) = \left\{
   \begin{array}{ll}
     h_0, & x \geq 0, \\
     0, & x < 0.
   \end{array}
  \right. \qquad
  u (x, 0) \equiv 0, \; \forall x\in\R.
\end{equation*}
Schematically it is depicted on Figure~\ref{fig:initcond}.

Then, by considering the Riemann invariants and using the method of characteristics, \cite{Lax1973, Godlewski1990, Godlewski1996, Chechkin2009}, one can derive the following solution:
\begin{equation}\label{eq:Stoker}
  H(x,t) = \left\{
   \begin{array}{lc}
     0, & x < -2c_0t, \\
     \frac{1}{9g}\Bigl(\frac{x}{t} + 2c_0\Bigr)^2, & -2c_0t \leq x \leq c_0t, \\
     h_0, & x > c_0t,
   \end{array}
  \right.
\end{equation}
\begin{equation}\label{eq:StokerU}
  u(x,t) = \left\{
   \begin{array}{lc}
     0, & x < -2c_0t, \\
     \frac23\Bigl(\frac{x}{t} - c_0\Bigr), & -2c_0t \leq x \leq c_0t, \\
     0, & x > c_0t,
   \end{array}
  \right.
\end{equation}
where $c_0 := \sqrt{gh_0}$ is the gravity wave speed in the undisturbed region. The front position is given by the characteristic outgoing from the fluid region:
\begin{equation*}
  x_f (t) = -2c_0 t.
\end{equation*}

Recall that recently this solution was generalized to the constant slope case by Mangeney {\sl et al}. (2000), \cite{Mangeney2000}.

\begin{remark}
The run-up algorithm used in our numerical code \VOLNA is based on this analytical result. Namely, we impose just obtained front speed when the wet/dry transition is detected. This simple approach was validated and shown to be very robust. For more details we refer to Dutykh {\sl et al}. (2009), \cite{Dutykh2009a}.
\end{remark}

%%%%%%%%%%%%%%%%%%%%%%%%%%%%%%%%%%%%%%%%%%%%%%%%%%%%%%%%%%%

\section{Numerical methods}\label{sec:num}

The main purpose of this study is to draw out some conclusions on the validity of \ac{NSWE} for wetting (flooding) process simulations. That is why, we do not provide here any details about numerical methods used to compute solutions. The interested reader can consult references given below to get technical details.

In order to solve numerically the two-fluid Navier-Stokes equations (\ref{eq:mass_pm}), (\ref{eq:momentum}) and (\ref{eq:advect}), we applied the finite volumes method, cf. e.g. \cite{Jasak1996, Rusche2002, OpenFOAM2007}. Namely, a freely available solver \texttt{interDyMFoam} of the \OpenFOAM CFD Toolbox \cite{OpenFOAM2007} was used. The interface between two fluids is reconstructed from the volume fraction $\phi$ distribution using the VOF method, cf. \cite{Hirt1981, Scardovelli1999, Popinet1999}. Let us underline that all two-fluid computations presented in this study are 3D with only one cell in $z$-direction. Everywhere we impose the classical no-slip boundary condition.

\acl{NSWE} are solved with our operational numerical code \VOLNA, cf. \cite{Dutykh2007a, Dutykh2009a}. This code was developed in close collaboration with R.~Poncet and F.~Dias when the first author was at CMLA, ENS de Cachan. The \VOLNA code uses unstructured triangular meshes and is able to run in arbitrary complex coastal regions. The numerical method is a second-order finite volumes MUSCL-TVD scheme along with the SSP-RK3(4) method for the discretization in time, cf. \cite{Spiteri2002}. Details on adopted discretization procedure can be found in \cite{Dutykh2007a, Poncet2008, Dutykh2009a}. All the computations we performed are in 2D and only one-dimensional cross sections are presented below. On the lateral boundaries we impose the wall boundary condition $\u\cdot\n = 0$. This choice is consistent with two-fluid computation and allows us to have an insight into the impact process.

\section{Comparison results and discussion}\label{sec:compare}

\begin{figure}
    \scalebox{0.9} % Change this value to rescale the drawing.
    {
\begin{pspicture}(0,-2.1029687)(13.582812,2.1029687)
\definecolor{color158g}{rgb}{0.8,0.8,1.0}
\definecolor{color158f}{rgb}{0.4,0.8,1.0}
\definecolor{color165f}{rgb}{0.0,0.2,0.8}
\psline[linewidth=0.04cm,arrowsize=0.05291667cm 2.0,arrowlength=1.4,arrowinset=0.4]{->}(6.4409375,-1.5554688)(6.4409375,2.0245314)
\psline[linewidth=0.04cm,arrowsize=0.05291667cm 2.0,arrowlength=1.4,arrowinset=0.4]{->}(6.4409375,-1.5554688)(13.500937,-1.5954688)
\psframe[linewidth=0.04,dimen=outer,fillstyle=gradient,gradlines=2000,gradbegin=color158g,gradend=color158f,gradmidpoint=1.0](12.440937,0.42453125)(0.5809375,-1.5754688)
\usefont{T1}{ptm}{m}{n}
\rput(13.282344,-1.9254688){$x$}
\usefont{T1}{ptm}{m}{n}
\rput(6.012344,1.9145312){$y$}
\psframe[linewidth=0.04,dimen=outer,fillstyle=gradient,gradlines=2000,gradend=color165f,gradmidpoint=1.0](12.440937,-0.55546874)(6.4009376,-1.5554688)
\usefont{T1}{ptm}{m}{n}
\rput(12.372344,-1.9054687){$\ell$}
\usefont{T1}{ptm}{m}{n}
\rput(6.3723435,-1.8854687){$0$}
\usefont{T1}{ptm}{m}{n}
\rput(0.65234375,-1.8654687){$-\ell$}
\usefont{T1}{ptm}{m}{n}
\rput(12.972343,-1.0854688){$h_0$}
\psline[linewidth=0.04cm,arrowsize=0.05291667cm 2.0,arrowlength=1.4,arrowinset=0.4]{<->}(12.620937,-0.59546876)(12.620937,-1.5354687)
\psline[linewidth=0.04cm,arrowsize=0.05291667cm 2.0,arrowlength=1.4,arrowinset=0.4]{<->}(0.4209375,0.40453124)(0.4209375,-1.5554688)
\usefont{T1}{ptm}{m}{n}
\rput(0.15,-0.68){$H$}
\end{pspicture}
    }
    \caption{Sketch of the initial condition for two-fluid numerical simulation (\ac{DNS}).}
    \label{fig:icond2phase}
\end{figure}
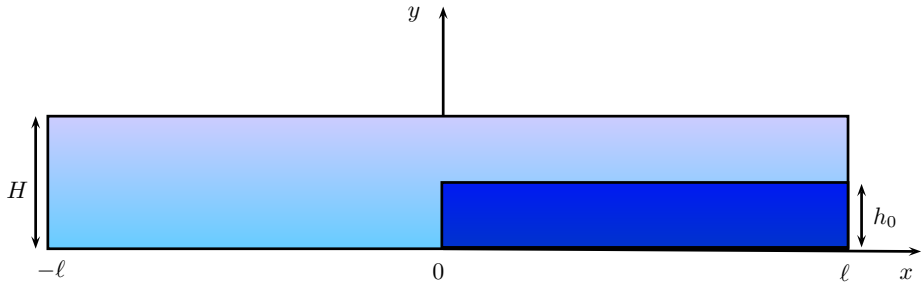

In this section we perform a comparison between a two-fluid simulation (\ac{DNS}), the analytical solution by Stoker (1957) and numerical solutions to the \ac{NSWE} by the \VOLNA code. The initial set-up for the \VOLNA code is shown in Figure \ref{fig:initcond}. Sketch of the initial condition for the \ac{DNS} is depicted on Figure~\ref{fig:icond2phase}. The simulation time and propagation distance is chosen so that the right boundary do not influence obtained results. All parameters used in computations are given in Table \ref{tab:params}. These parameters are chosen suitable to simulate the air/water interaction.

\begin{table}
  \begin{center}
    \begin{tabular}{c|l}
      \textit{parameter} & \textit{value} \\
      \hline\hline
      gravity acceleration, $g$, $m/s^2$ & $1.0$ \\
      \hline
      fluid column height, $h_0$, $m$ & $0.25$ \\
      \hline
      fluid column length, $\ell$, $m$ & $1.0$ \\
      \hline
      total domain height, $H$, $m$ & $0.5$ \\
      \hline
      fluid density, $\rho^+$, $kg/m^3$ & $1000.0$ \\
      \hline
      fluid viscosity, $\nu^+$, $m^2/s$ & $10^{-6}$ \\
      \hline
      air density, $\rho^-$, $kg/m^3$ & $1.0$ \\
      \hline
      air viscosity, $\nu^-$, $kg/m^3$ & $10^{-6}$ \\
      \hline
      surface tension, $\sigma$, $kg$ & $0.07$ \\
      %\hline\hline
    \end{tabular}
    \caption{Parameters used in numerical simulations.}
    \label{tab:params}
  \end{center}
\end{table}

\begin{figure}
  \centering
  \subfigure[Two-fluid simulation]%
  {\includegraphics[width=0.6\textwidth]{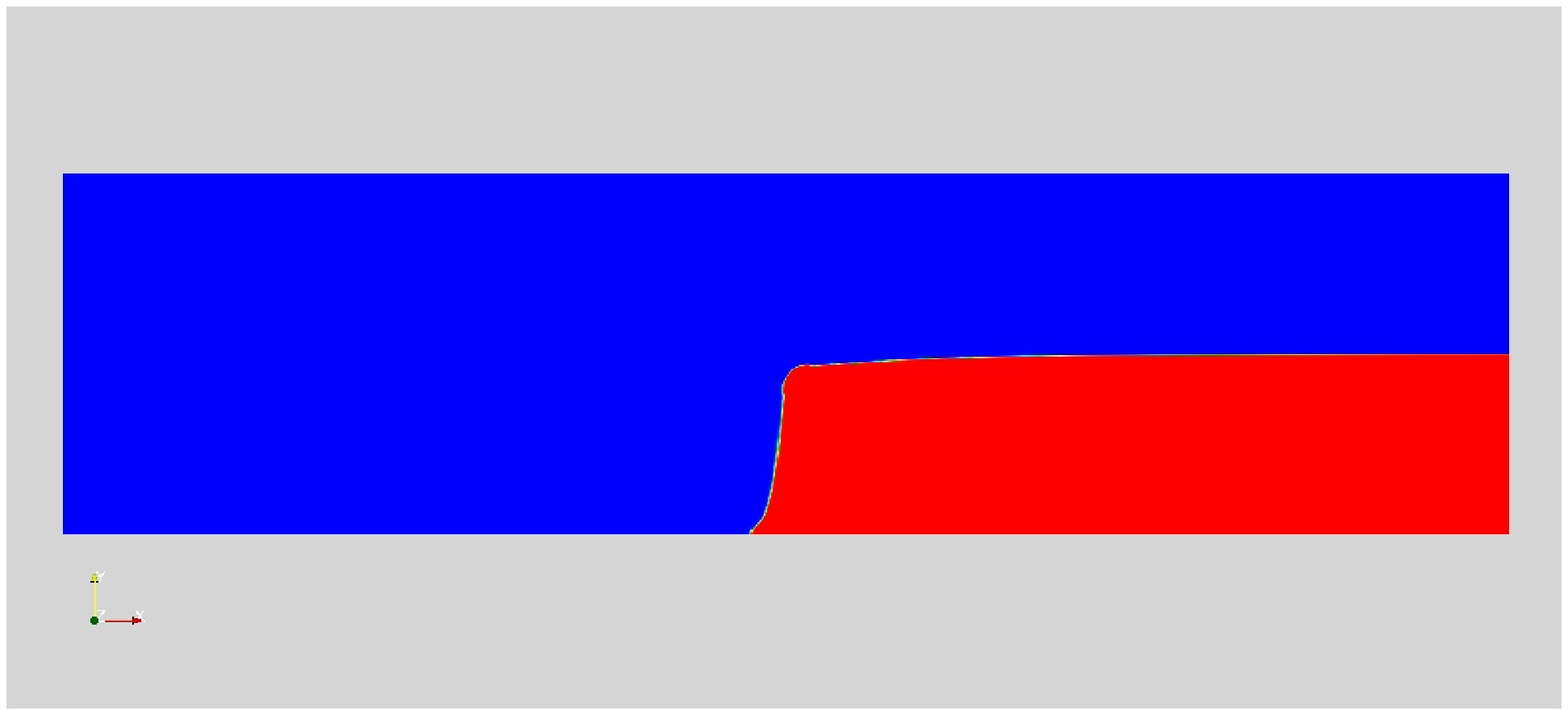}}
  \subfigure[\ac{NSWE}]%
  {\includegraphics[width=0.39\textwidth]{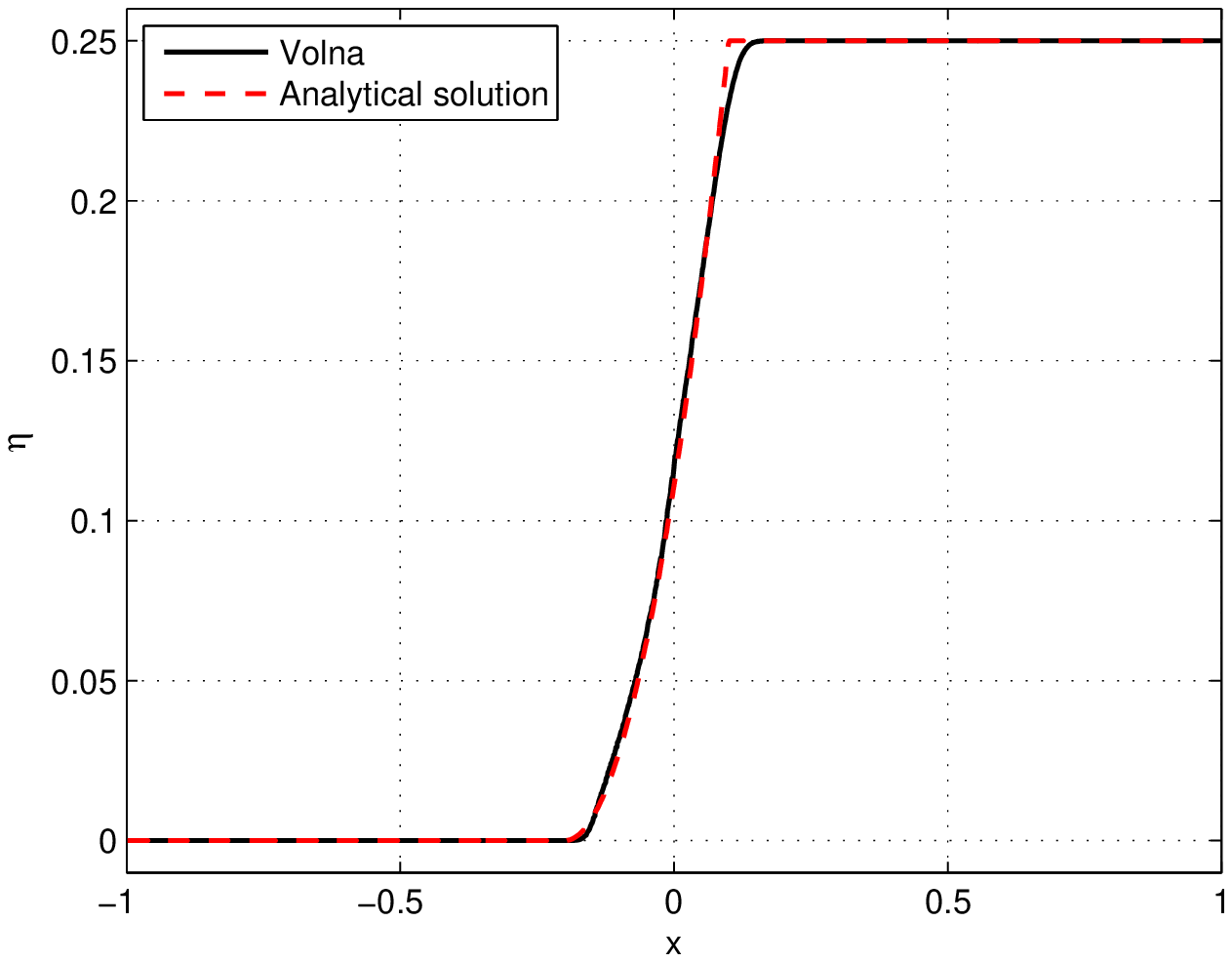}}
  \caption{Initial deformation of the water column under the gravity force ($t = 0.2$ s).}
  \label{fig:slidingBegin}
\end{figure}

\begin{figure}
  \centering
  \subfigure[Two-fluid simulation]%
  {\includegraphics[width=0.6\textwidth]{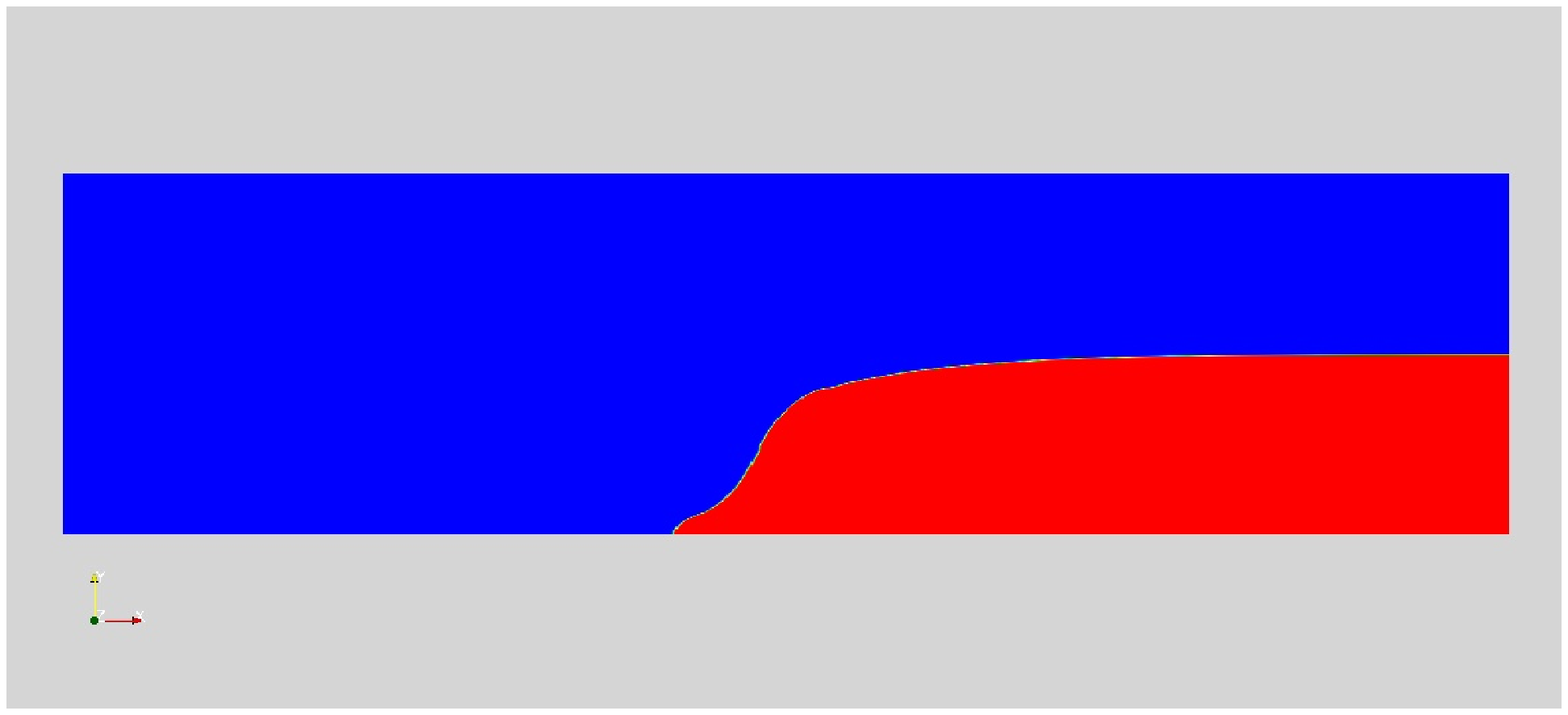}}
  \subfigure[\ac{NSWE}]%
  {\includegraphics[width=0.39\textwidth]{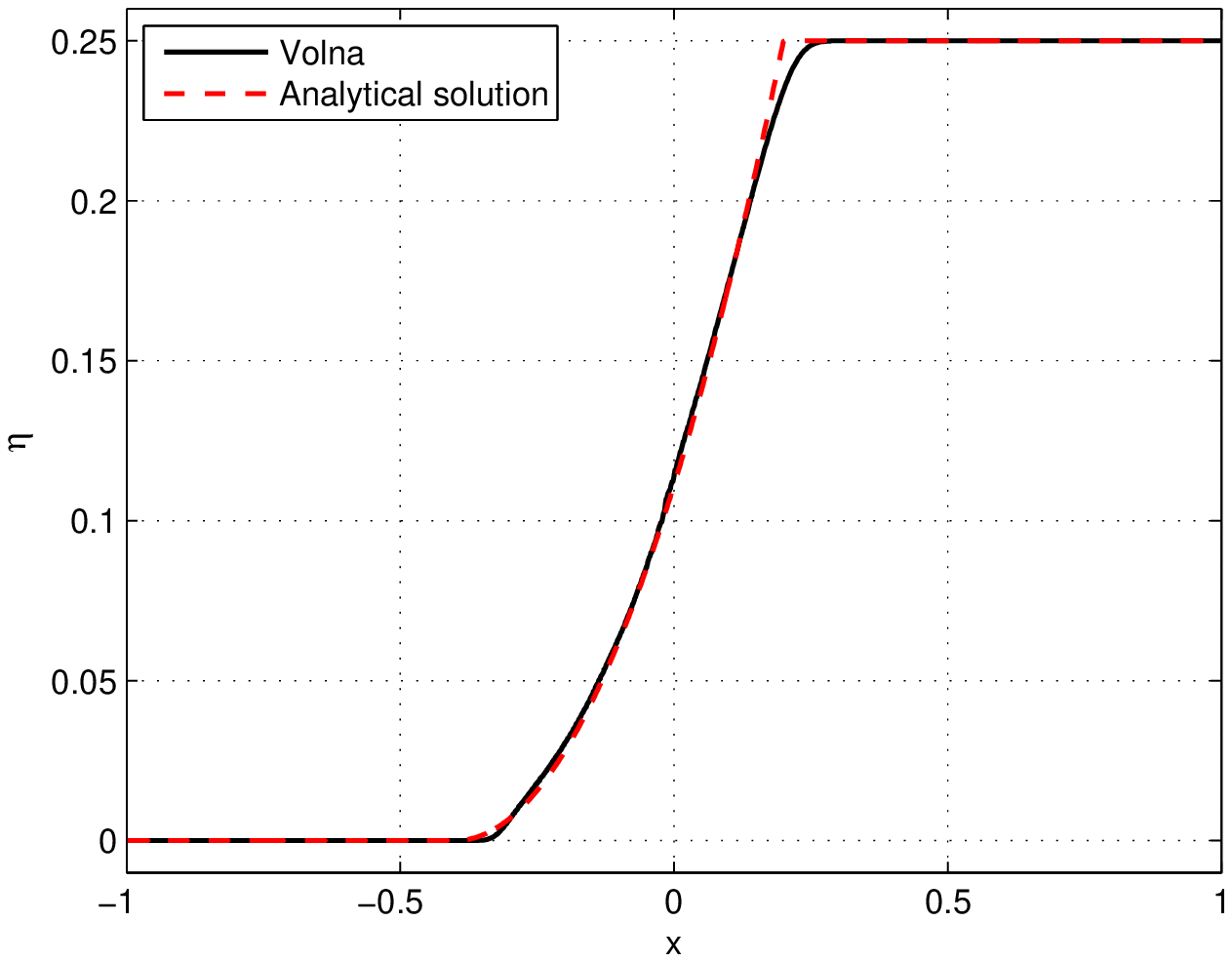}}
  \caption{Transition to the propagation r\'egime ($t = 0.4$ s).}
  \label{fig:slidingSecond}
\end{figure}

\begin{figure}
  \centering
  \subfigure[Two-fluid simulation]%
  {\includegraphics[width=0.6\textwidth]{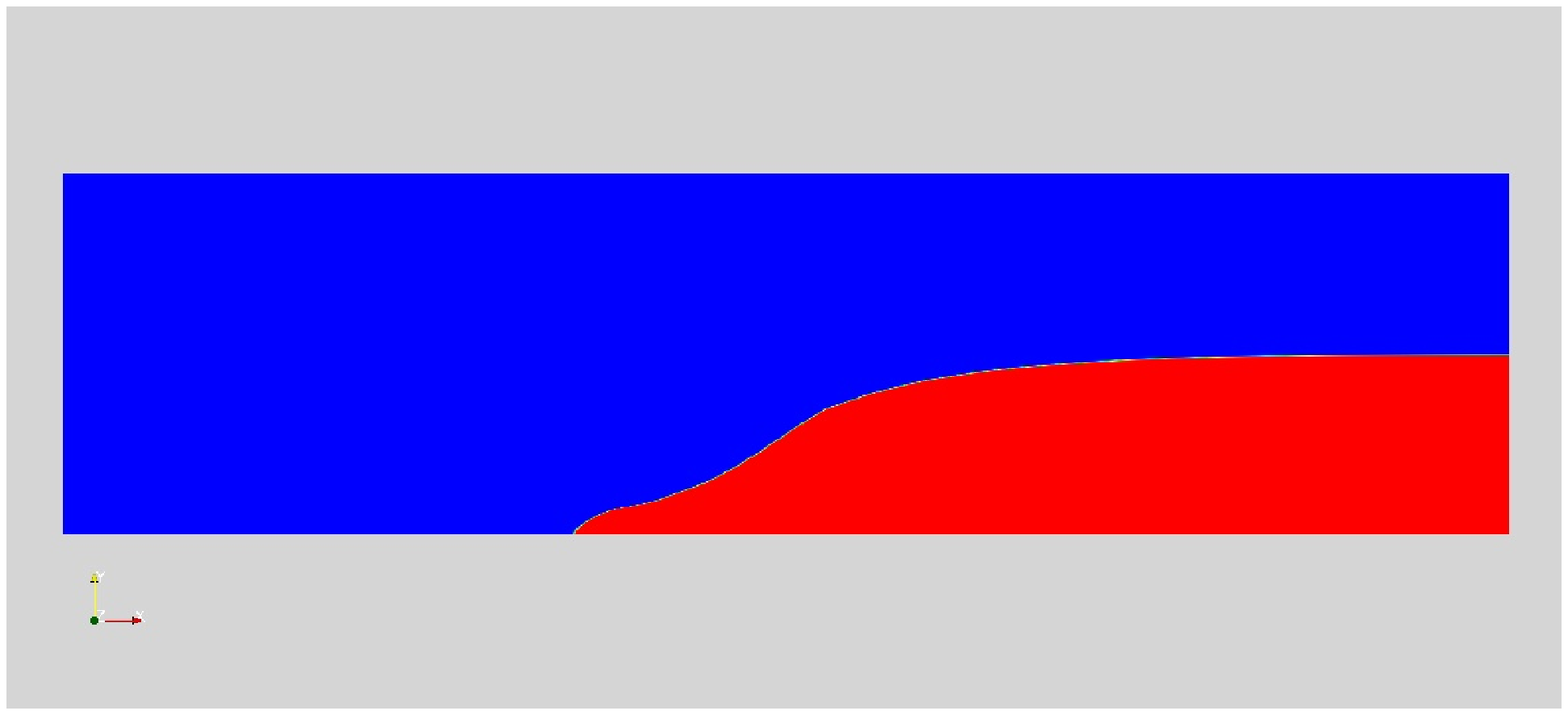}}
  \subfigure[\ac{NSWE}]%
  {\includegraphics[width=0.39\textwidth]{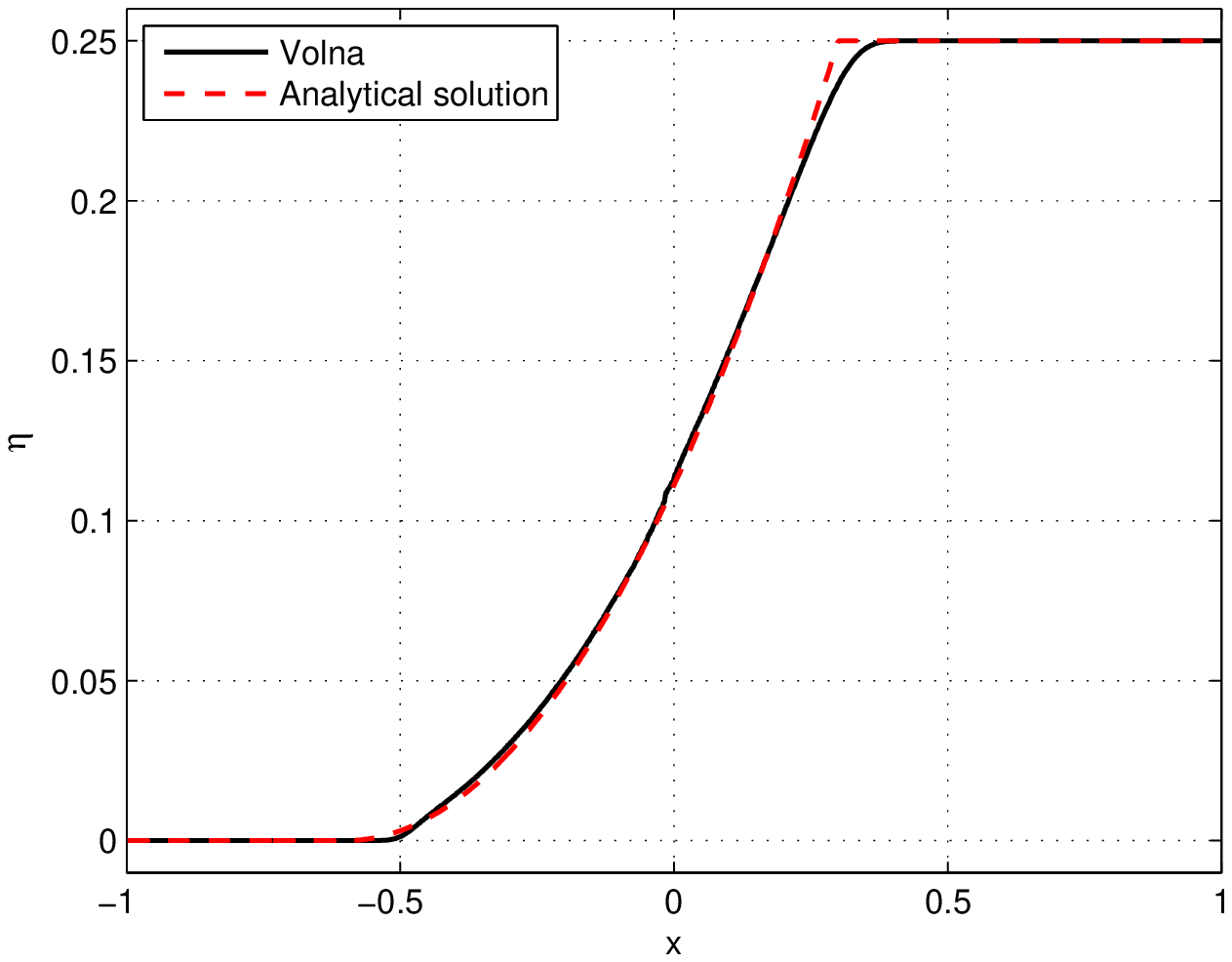}}
  \caption{Heavy fluid entering into the propagation r\'egime ($t = 0.6$ s).}
  %\label{fig:slidingEnd}
\end{figure}

\begin{figure}
  \centering
  \subfigure[Two-fluid simulation]%
  {\includegraphics[width=0.6\textwidth]{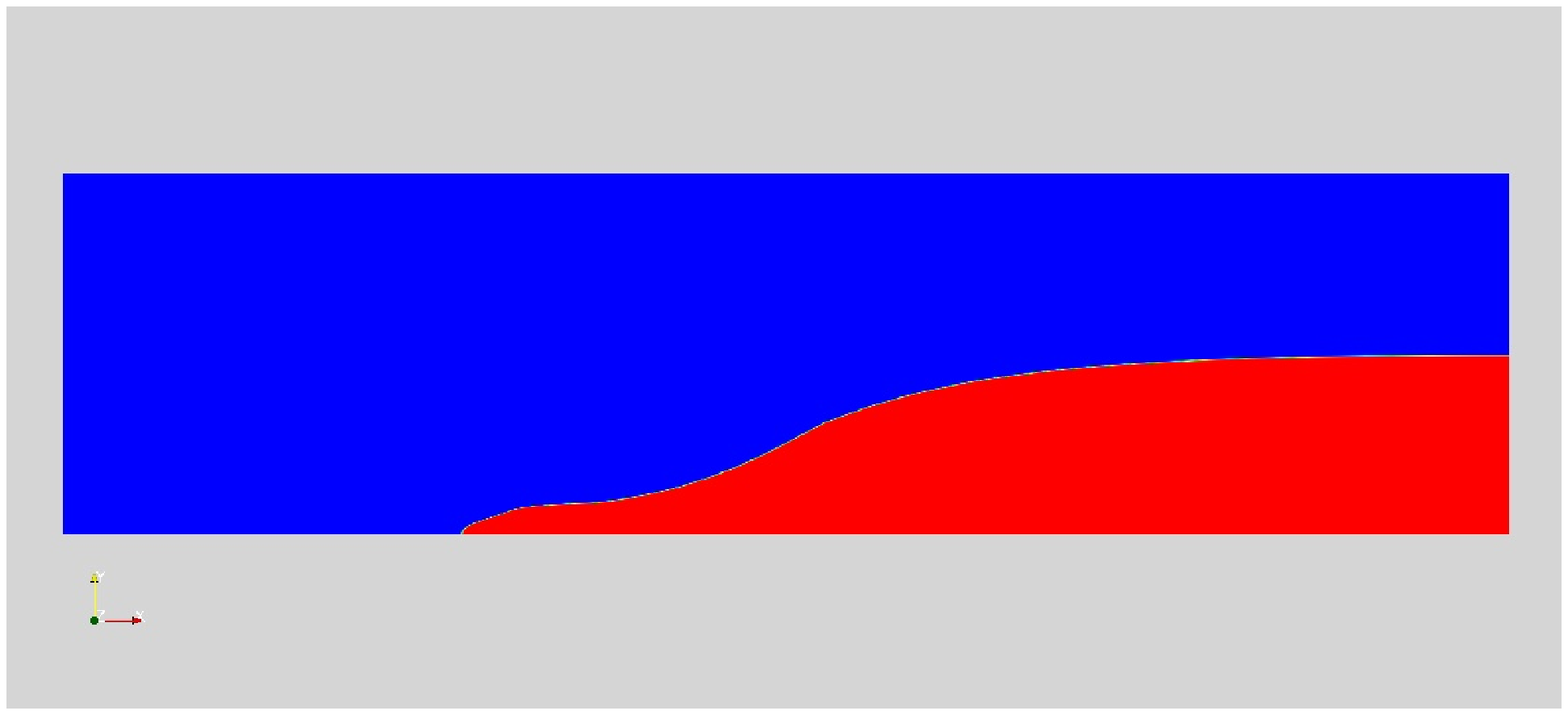}}
  \subfigure[\ac{NSWE}]%
  {\includegraphics[width=0.39\textwidth]{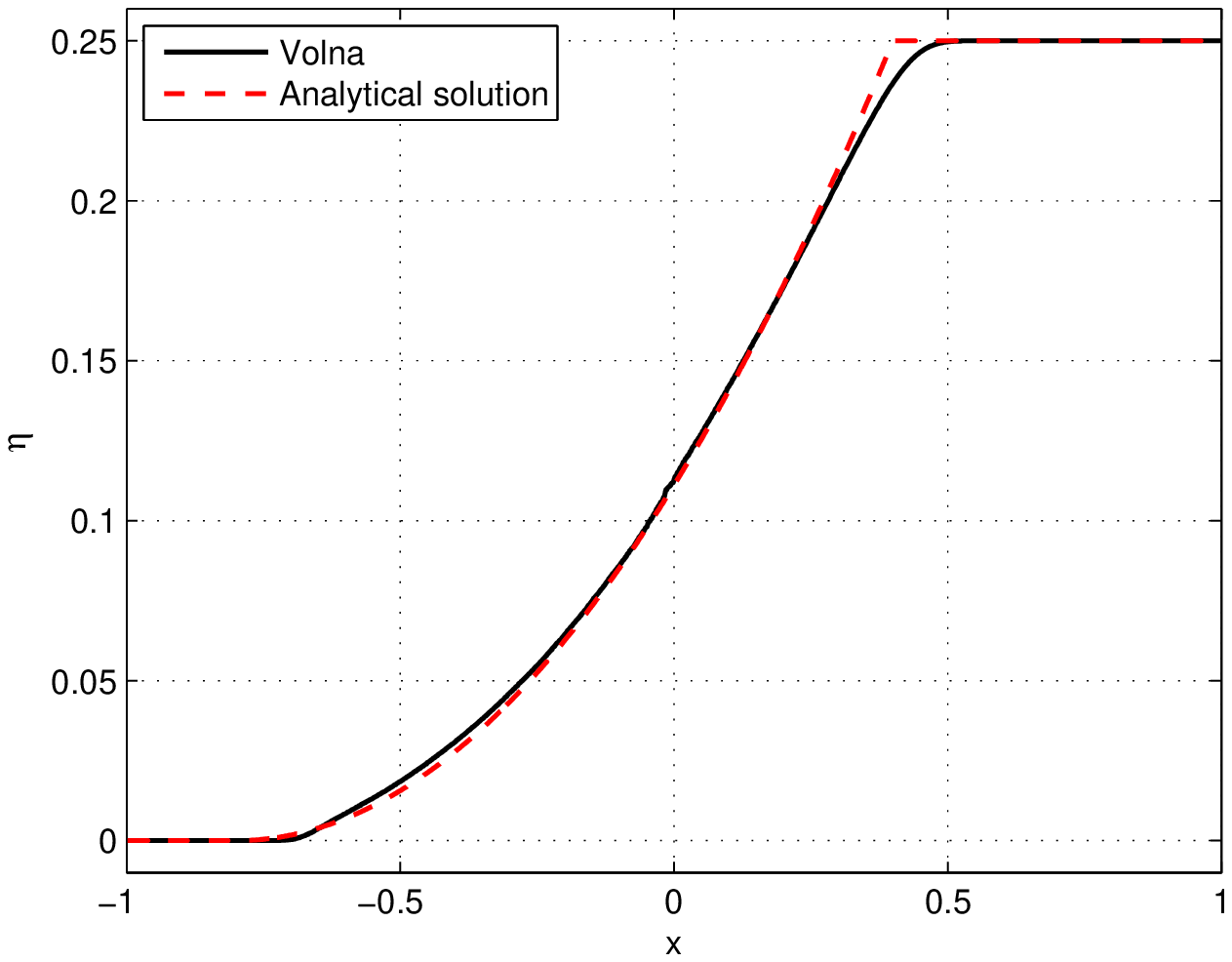}}
  \caption{Heavy fluid in the propagation r\'egime ($t = 0.8$ s).}
  %\label{fig:slidingEnd}
\end{figure}

\begin{figure}
  \centering
  \subfigure[Two-fluid simulation]%
  {\includegraphics[width=0.6\textwidth]{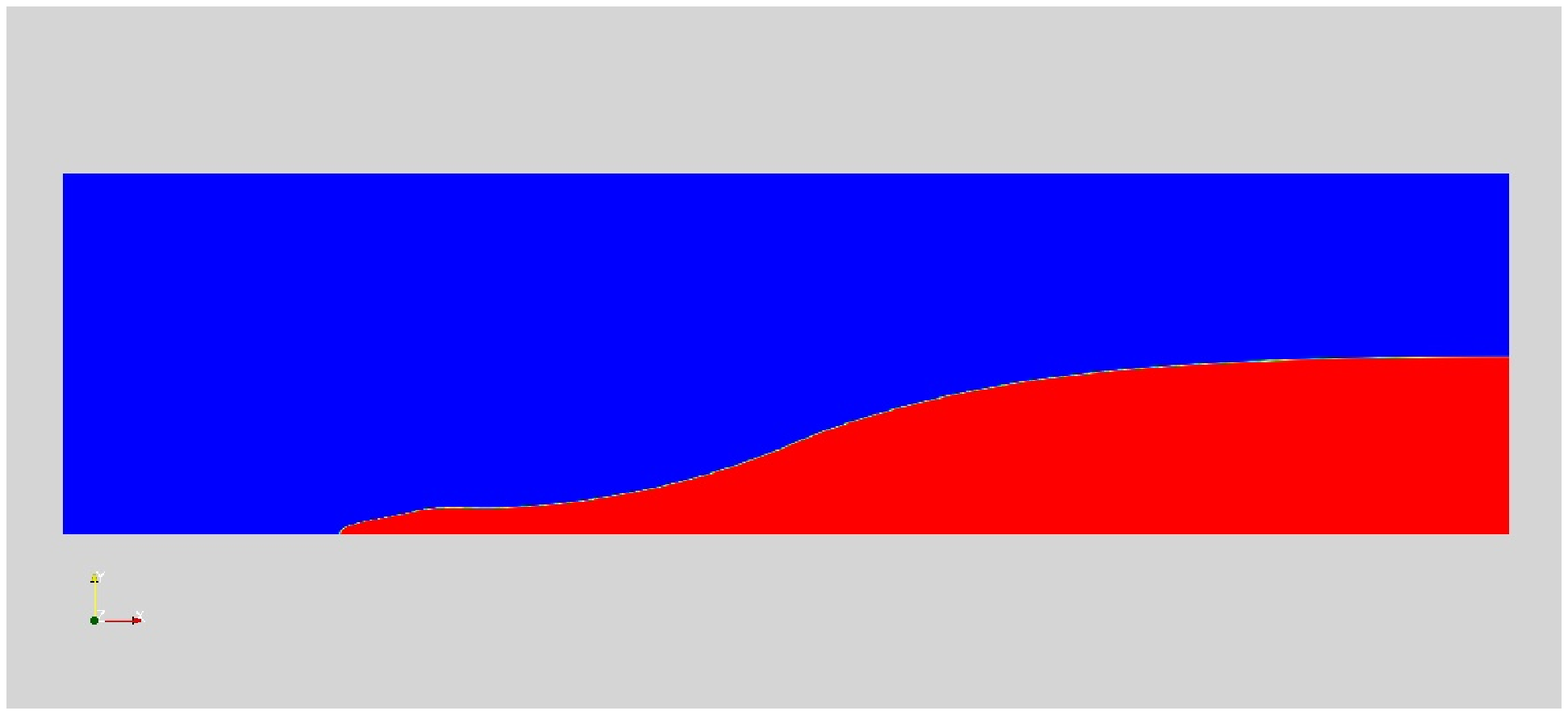}}
  \subfigure[\ac{NSWE}]%
  {\includegraphics[width=0.39\textwidth]{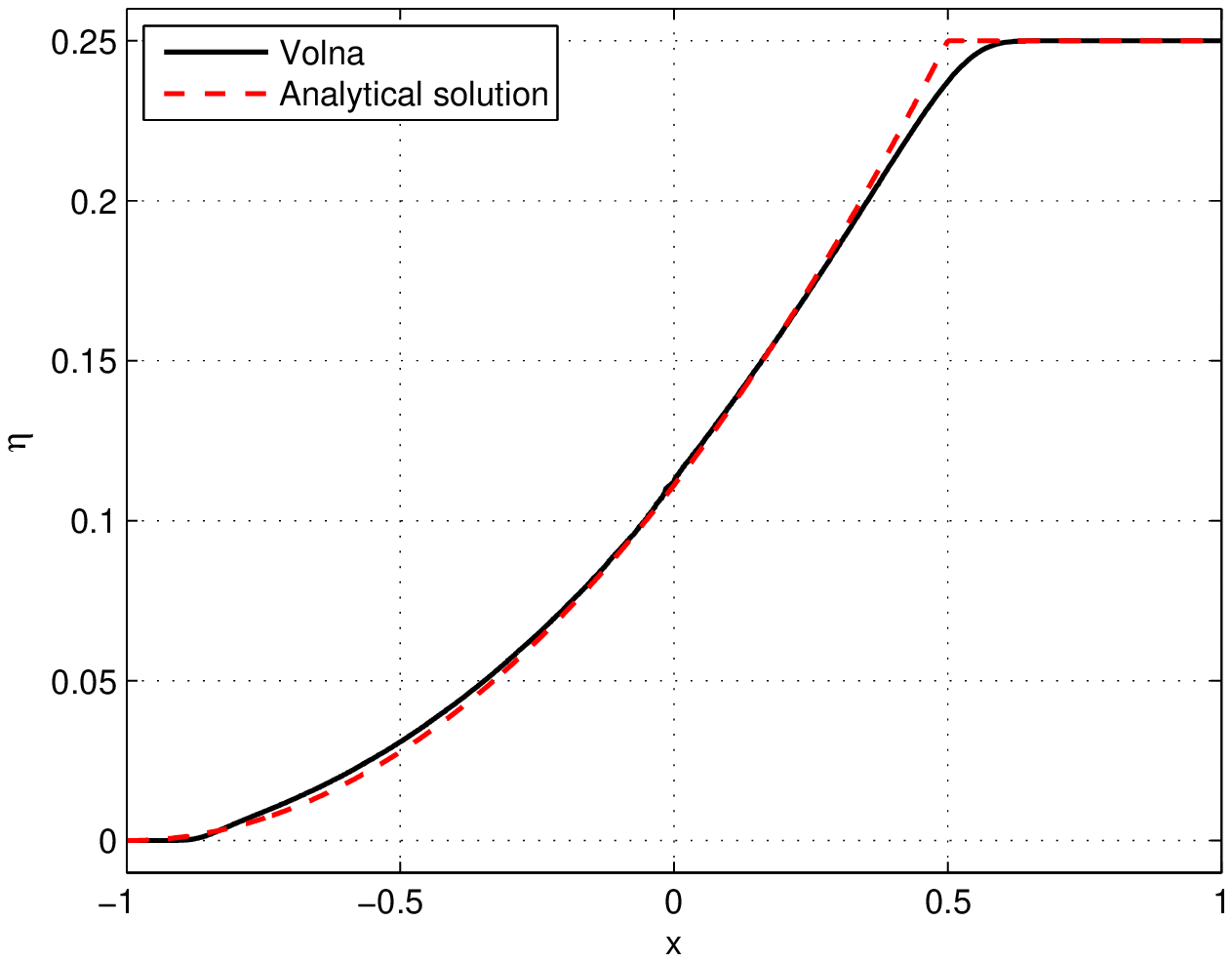}}
  \caption{Heavy fluid in the propagation r\'egime ($t = 1$ s).}
  %\label{fig:slidingEnd}
\end{figure}

\begin{figure}
  \centering
  \subfigure[Two-fluid simulation]%
  {\includegraphics[width=0.6\textwidth]{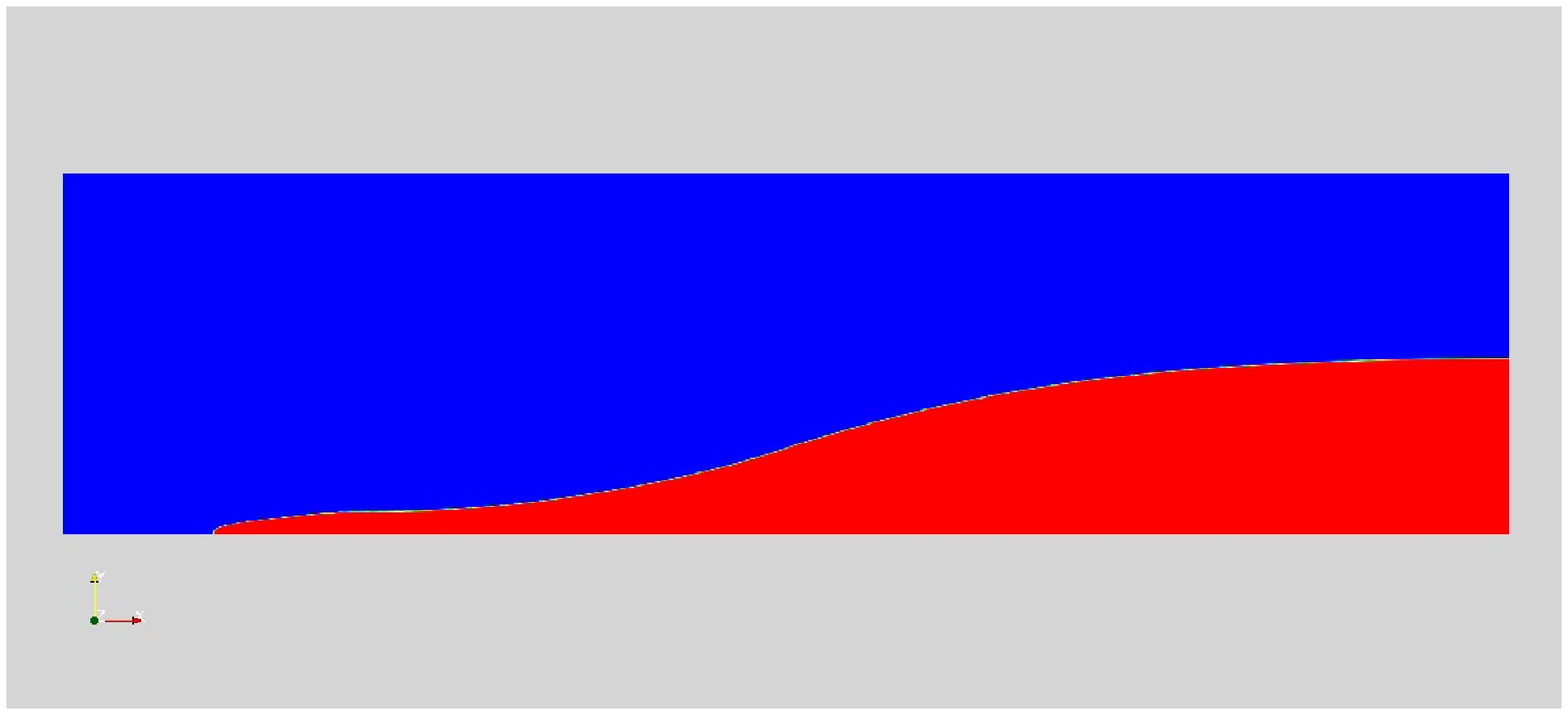}}
  \subfigure[\ac{NSWE}]%
  {\includegraphics[width=0.39\textwidth]{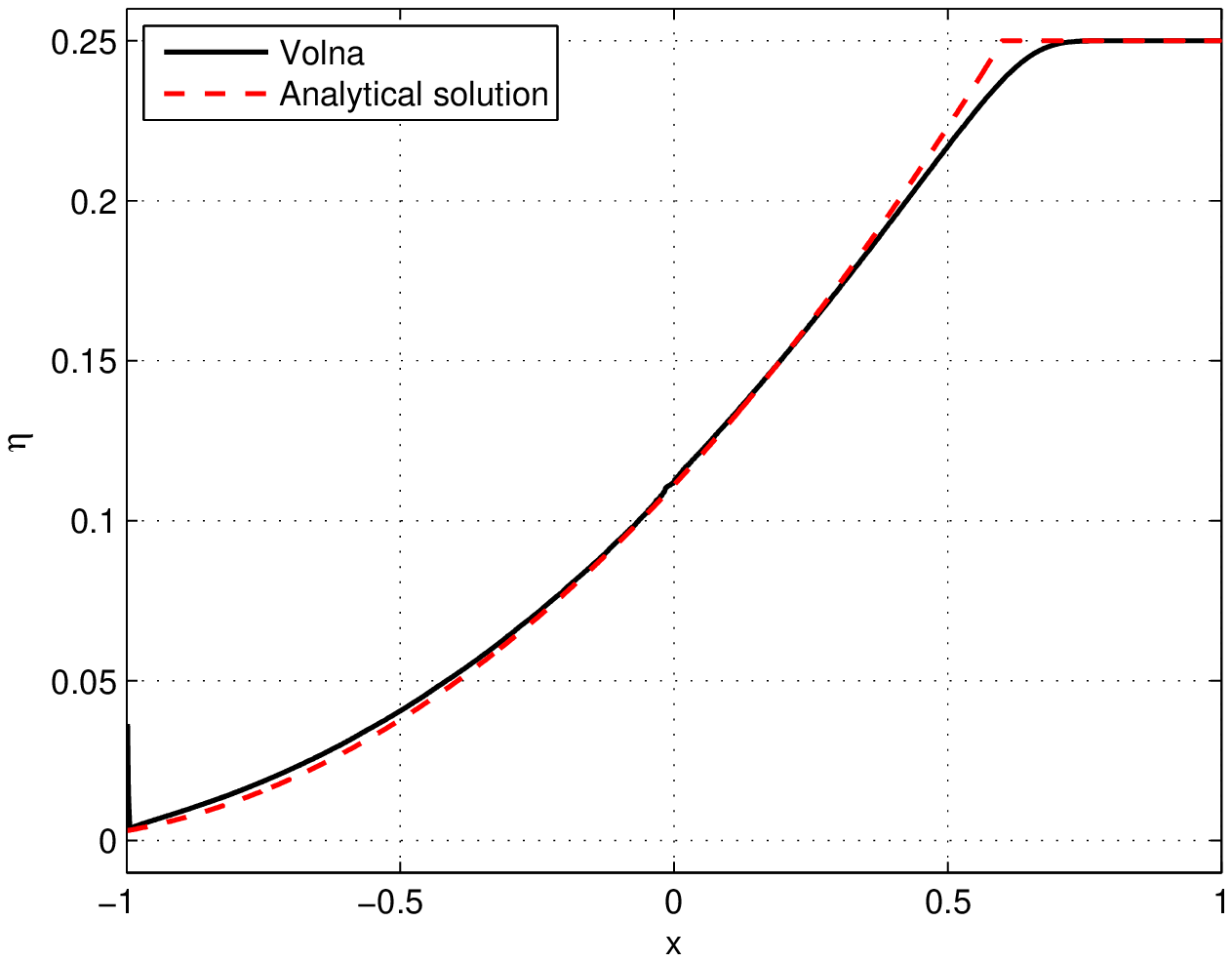}}
  \caption{Heavy fluid front before the interaction with the left wall ($t = 1.2$ s).}
  \label{fig:slidingAlmostEnd}
\end{figure}

\begin{figure}
  \centering
  \subfigure[Two-fluid simulation]%
  {\includegraphics[width=0.6\textwidth]{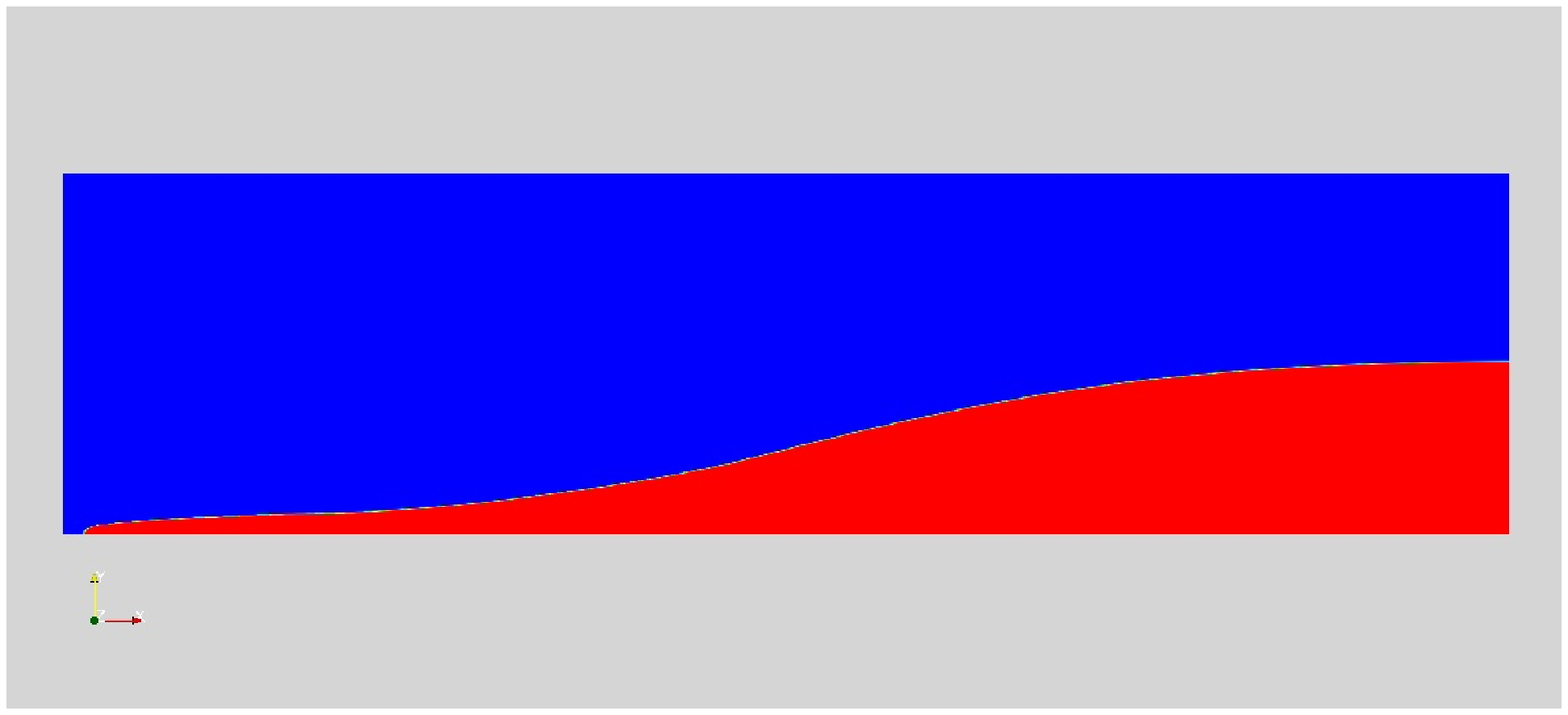}}
  \subfigure[\ac{NSWE}]%
  {\includegraphics[width=0.39\textwidth]{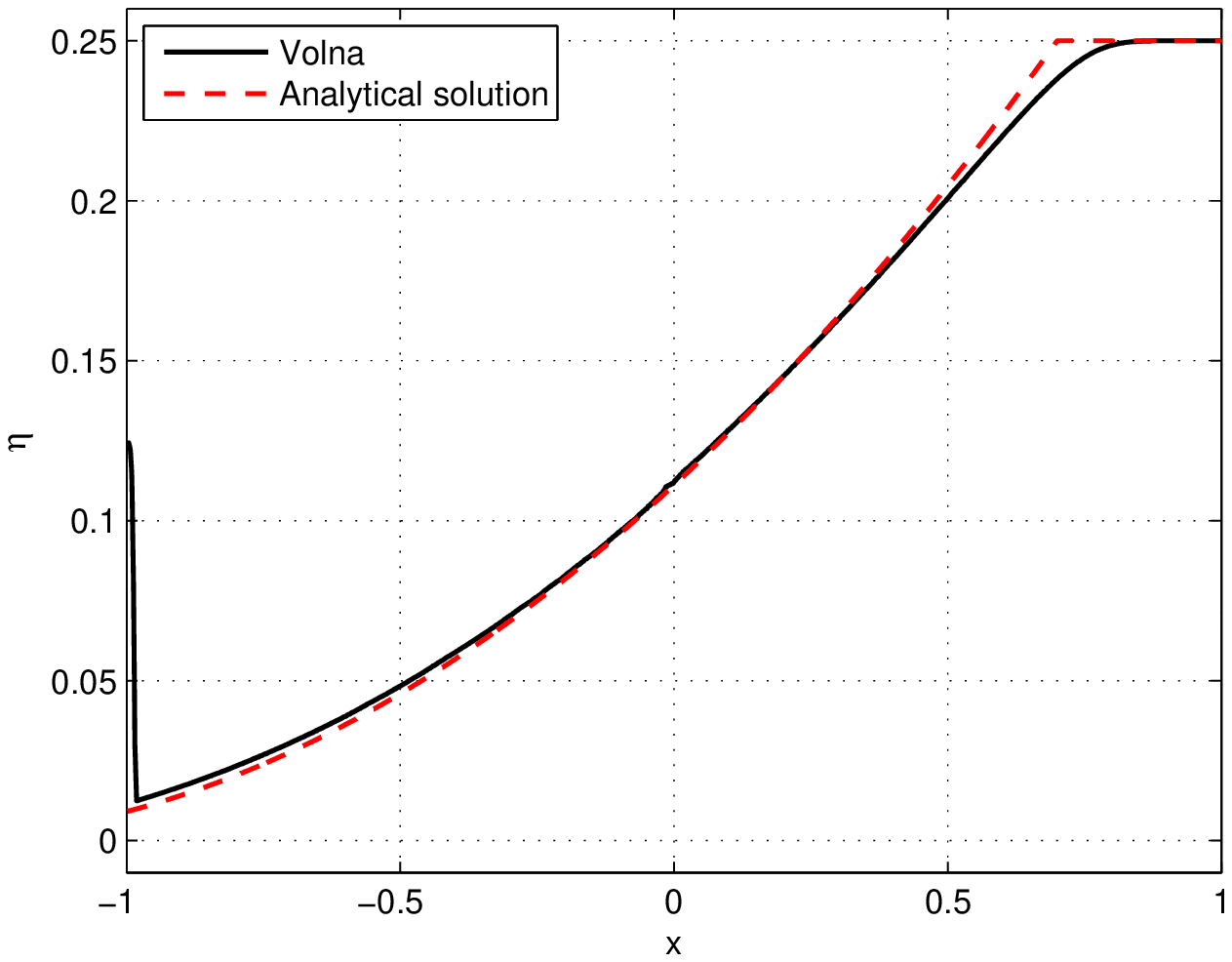}}
  \caption{Heavy fluid front before the interaction with the left wall ($t = 1.4$ s).}
  \label{fig:slidingEnd}
\end{figure}

The snapshots of our simulations are given on Figures \ref{fig:slidingBegin} -- \ref{fig:slidingEnd}. On the left image (a) we represent the volume fraction $\phi$ distribution provided by the \ac{DNS}. On the right image (b) we plot together the analytical solution (\ref{eq:Stoker}) (red dotted line) and simulation results by the \VOLNA code (black solid line) for the free surface elevation $\eta$. The analytical solution is almost superposed with our numerical simulation as it is expected. This result can be considered as one more validation test of the wetting/drying algorithm used in the \VOLNA code.

\begin{figure}
    \centering
        \includegraphics[width=0.99\textwidth]{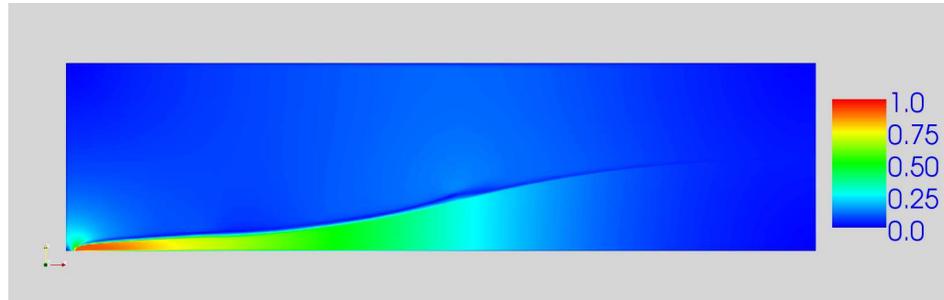}
    \caption{Velocity field magnitude at $t=1.4$ s.}
    \label{fig:u_mag1_4}
\end{figure}

In the beginning of the simulation, the water column is slightly deformed due to the gravity force (Figure \ref{fig:slidingBegin}). Only a small time interval is needed for the heavy fluid to acquire the kinetic energy and to enter into the propagation r\'egime depicted on Figures \ref{fig:slidingSecond} -- \ref{fig:slidingEnd}. Analytical solution (\ref{eq:Stoker}) prescribes a parabolic form of the interface. However, the \ac{DNS} shows somehow different shape. Lower fluid layers undergo stronger acceleration than in \ac{NSWE} and thus propagate faster. Nonuniform distribution of the velocity field along the heavy fluid is illustrated on Figure \ref{fig:u_mag1_4}. This creates a strong distortion of the interface which is elongated near the bottom (it can be easily seen in Figures \ref{fig:slidingAlmostEnd} -- \ref{fig:slidingEnd}). This effect is not present in \ac{NSWE} simulations since, the vertical flow structure is not resolved by this approximate model. Consequently, in \ac{NSWE} we obtain a piecewise linear distribution of the velocity field as it follows from analytical solution (\ref{eq:StokerU}).

%%%%%%%%%%%%%%%%%%%%%%%%%%%%%%%%%%%%%%%%%%%%

%%%%%%%%%%%%%%%%%%%%%%%%%%%%%%%%%%%%%%%%%%%%

Let us notice another one fact. In Figure \ref{fig:slidingEnd} one can observe that the \ac{NSWE} solution has already reached the left vertical wall. From this moment, the analytical solution is not valid anymore. However, the two-fluid simulation has not yet reached the left boundary. This discrepancy comes from the time lag due to initial acceleration stage, on one hand, and slightly different front propagation speeds, on the other hand. Bottom boundary layer may have some effect onto the propagation speed of the heavy fluid front \cite{Dutykh2007a, Dutykh2008a}.

%%% Front speed %%%

\begin{figure}
  \centering
  \subfigure[$h_0 = 0.25$ m]%
  {\includegraphics[width=0.49\textwidth]{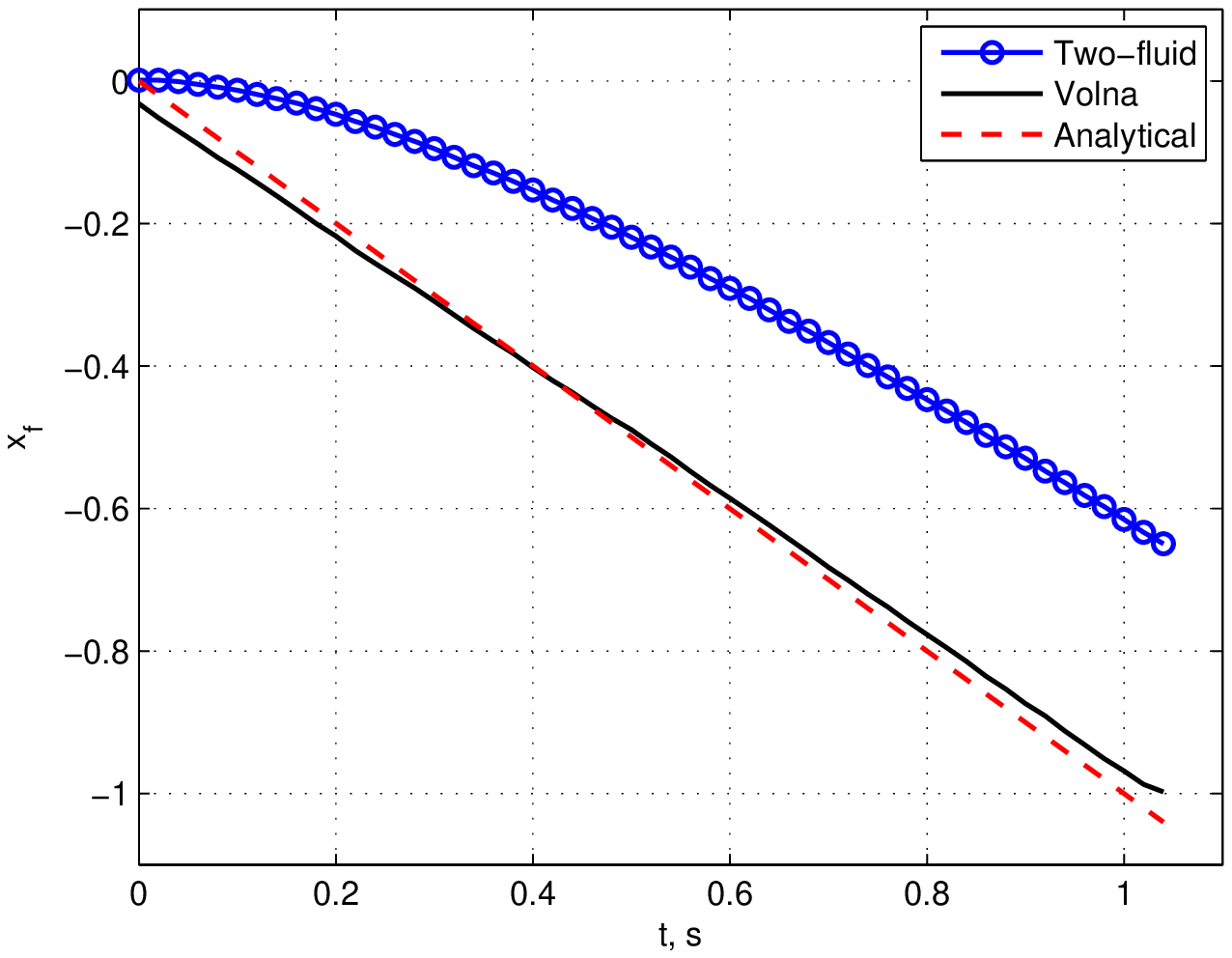}}
  \subfigure[$h_0 = 0.125$ m]%
  {\includegraphics[width=0.49\textwidth]{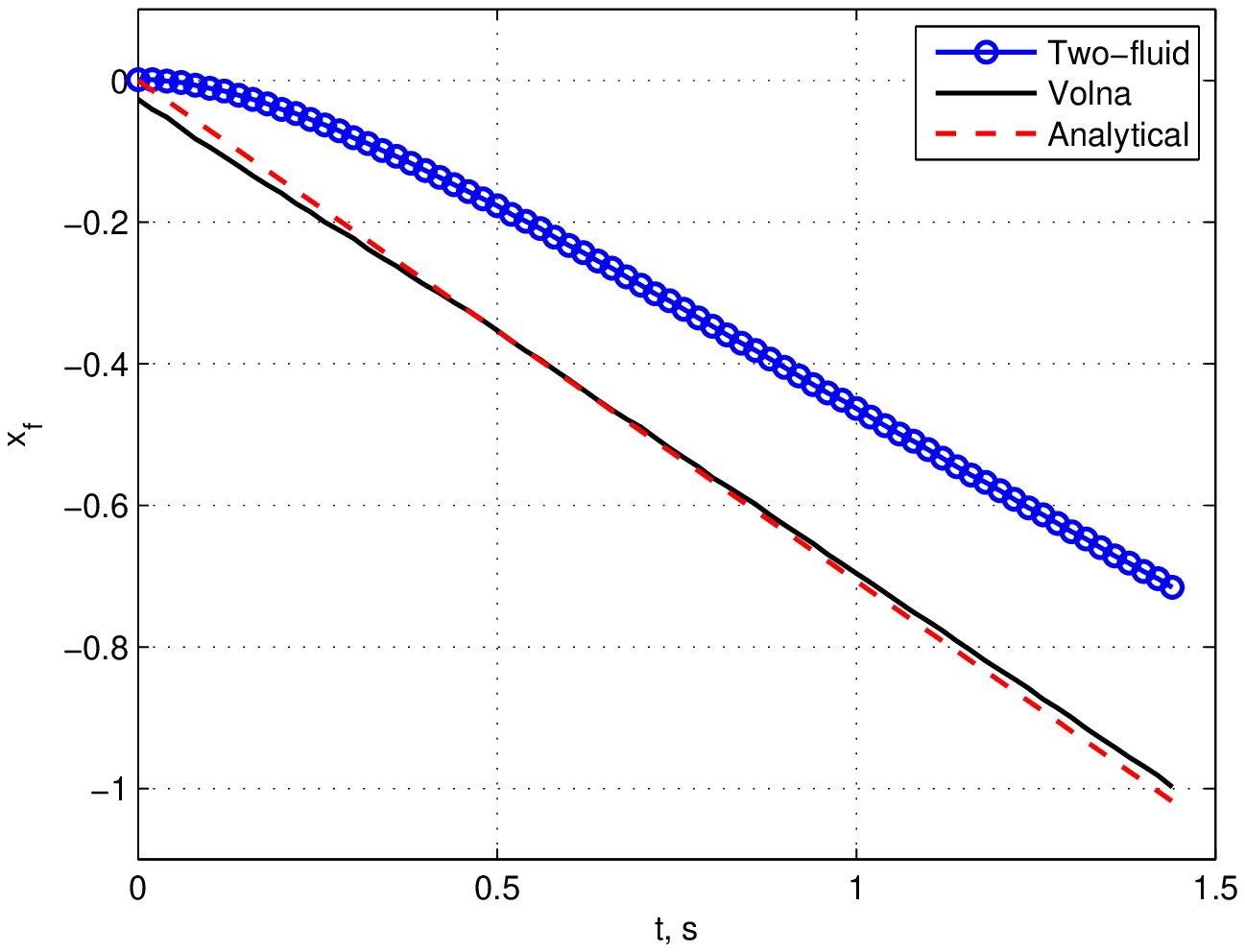}}
  \caption{Comparison of the front position according to three different models: blue line with circles corresponds to the \ac{DNS}, black line shows the front position predicted by \VOLNA solver (\ac{NSWE}) and the red dotted line is the analytical solution (\ref{eq:Stoker}).}
  \label{fig:front}
\end{figure}

From these simulations, we extracted the wave front position, shown in Figure \ref{fig:front} as a function of time for two different initial heights: $h_0 = 0.25$ m (as in simulations presented above) and $h_0 = 0.125$ m. Qualitatively these two results are similar. We can underline again a very good agreement between numerical and analytical \ac{NSWE} results. On the other hand, there is a slightly increasing difference in the front position with the two-fluid \ac{DNS}. It can be attributed to the initial acceleration stage ($t \leq 0.25$ s) which is present in the two-fluid model. Another explanation consists in the front velocity which can be determined by measuring the slope to the curve $t\to x_f(t)$. Determined in this way front speeds (in permanent r\'egime, $t\geq 0.25$ s) are given in Table \ref{tab:front}. The \ac{LSWE} give completely wrong results showing again that the nonlinearity plays the crucial r\^ole in this process. It is also worth to note that the numerical front speed by the \VOLNA code is closer to the \ac{DNS}. This positive fact can be attributed to the effect of the numerical diffusion on unstructured meshes.

\begin{table}
  \begin{center}
    \begin{tabular}{c||c|c}
          & $h_0 = 0.25$ m & $h_0 = 0.125$ m \\
        \hline\hline
        \textit{Two-fluid (\ac{DNS})} & $-0.81$ & $-0.58$ \\
        \hline
        \textit{\VOLNA (\ac{NSWE})} & $-0.96$ & $-0.68$ \\
        \hline
        \textit{Analytical (\ac{NSWE})} & $-1.0$ & $-0.71$ \\
        \hline
        \textit{Analytical (\ac{LSWE})} & $-0.5$ & $-0.35$ \\
    \end{tabular}
    \caption{Front speed predicted by four different approaches.}
    \label{tab:front}
  \end{center}
\end{table}

\subsection{Impact process}

In the previous section we presented results concerning the initial and propagation stages of the dam break problem. However, we continued the computations until the interaction with the left wall and even slightly beyond. The goal is to test again the validity of the \ac{NSWE} in such extreme conditions. For this kind of situations there is no analytical solution, and thus, we compare only \ac{DNS} and \VOLNA code results. In two-fluid simulation we use the same no-slip condition for all solid boundaries. The implementation of the wall boundary condition of \VOLNA solver can be found in \cite{Dutykh2009a}, and it is based on considerations of incoming characteristics. The general methodology is presented in works of J.-M. Ghidaglia and F. Pascal \cite{Ghidaglia2002a, Ghidaglia2002, Ghidaglia2005}.

\begin{figure}
  \centering
  \subfigure[Two-fluid simulation]%
  {\includegraphics[width=0.6\textwidth]{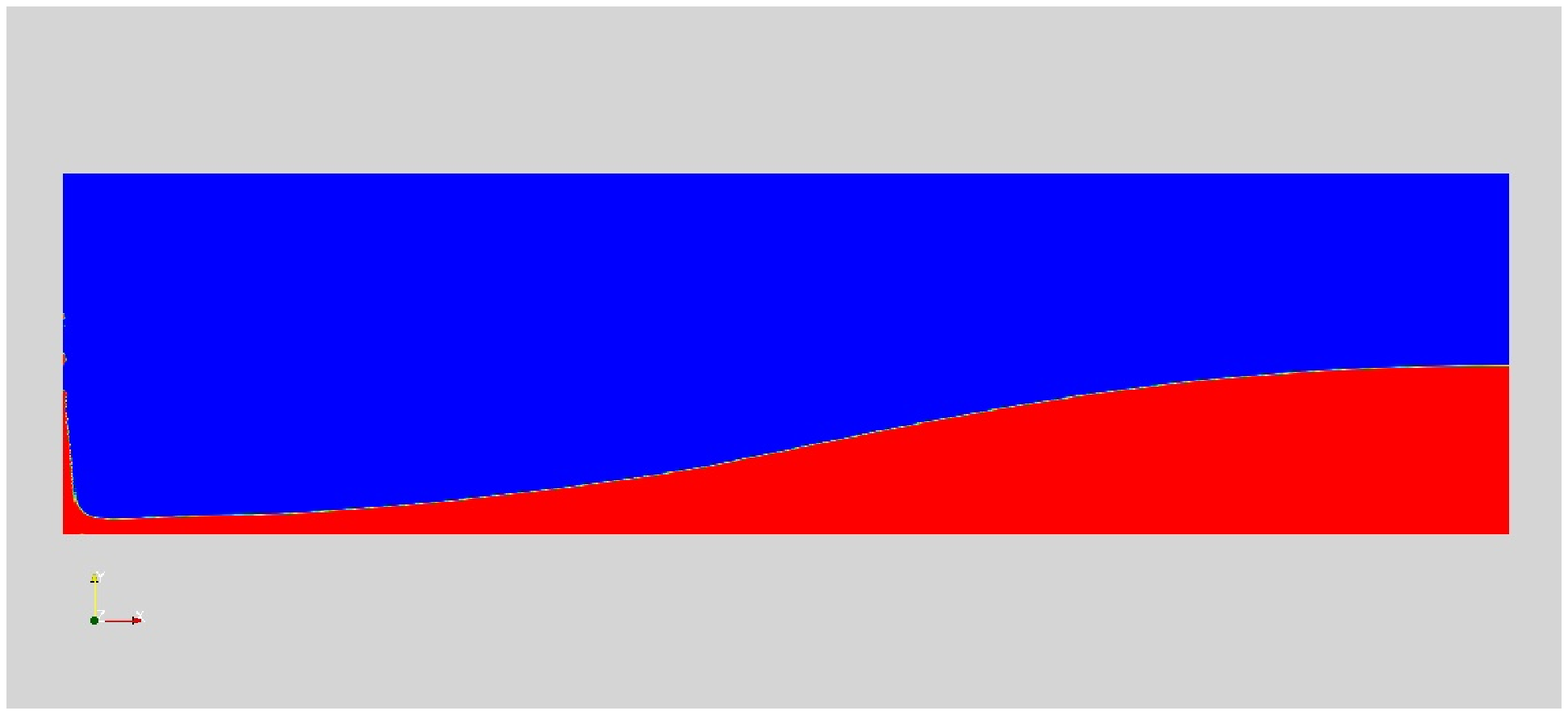}}
  \subfigure[\ac{NSWE} with \VOLNA solver]%
  {\includegraphics[width=0.39\textwidth]{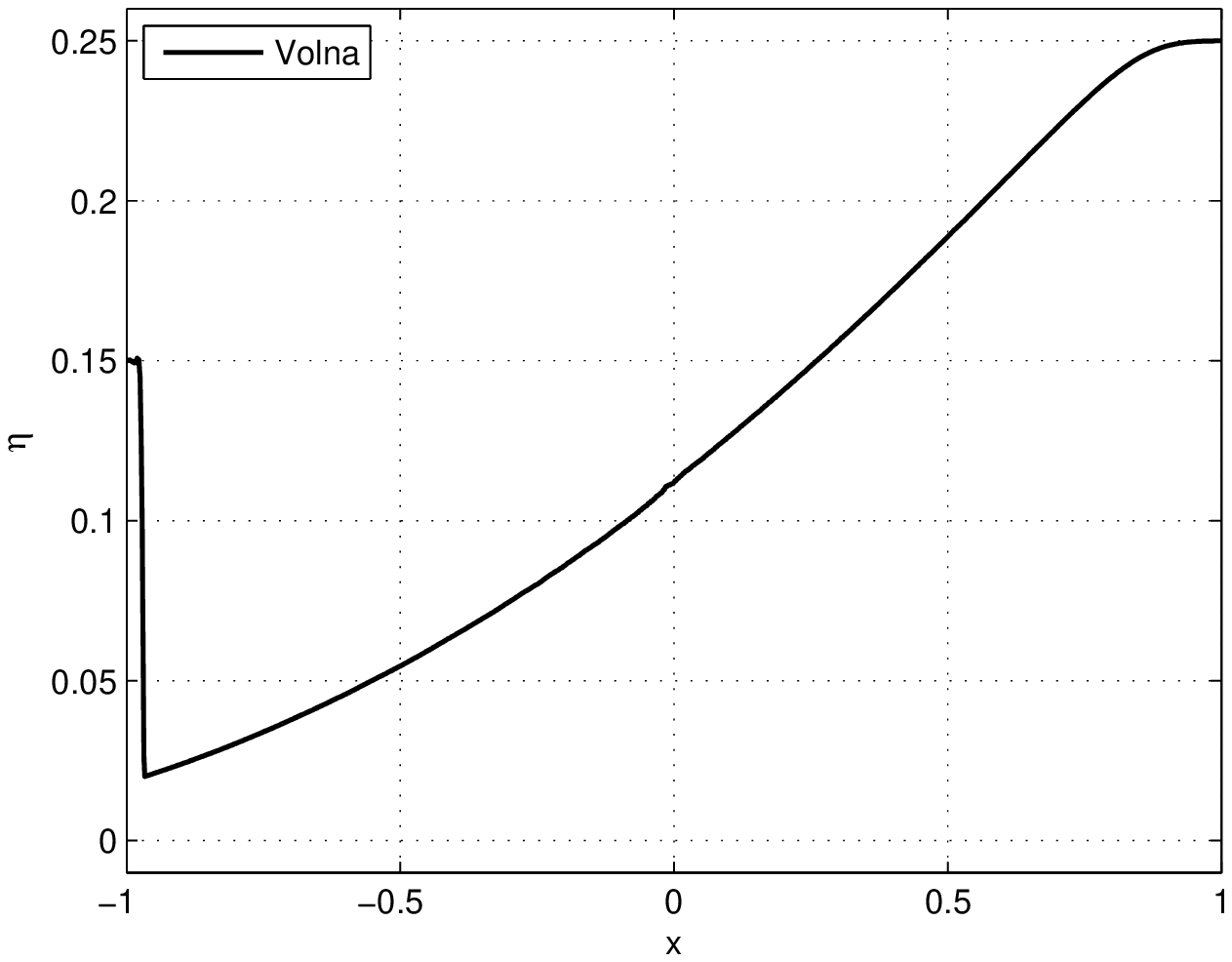}}
  \caption{Interaction with the left vertical wall at $t = 1.6$ s.}
  \label{fig:impact1}
\end{figure}

\begin{figure}
  \centering
  \subfigure[Two-fluid simulation]%
  {\includegraphics[width=0.6\textwidth]{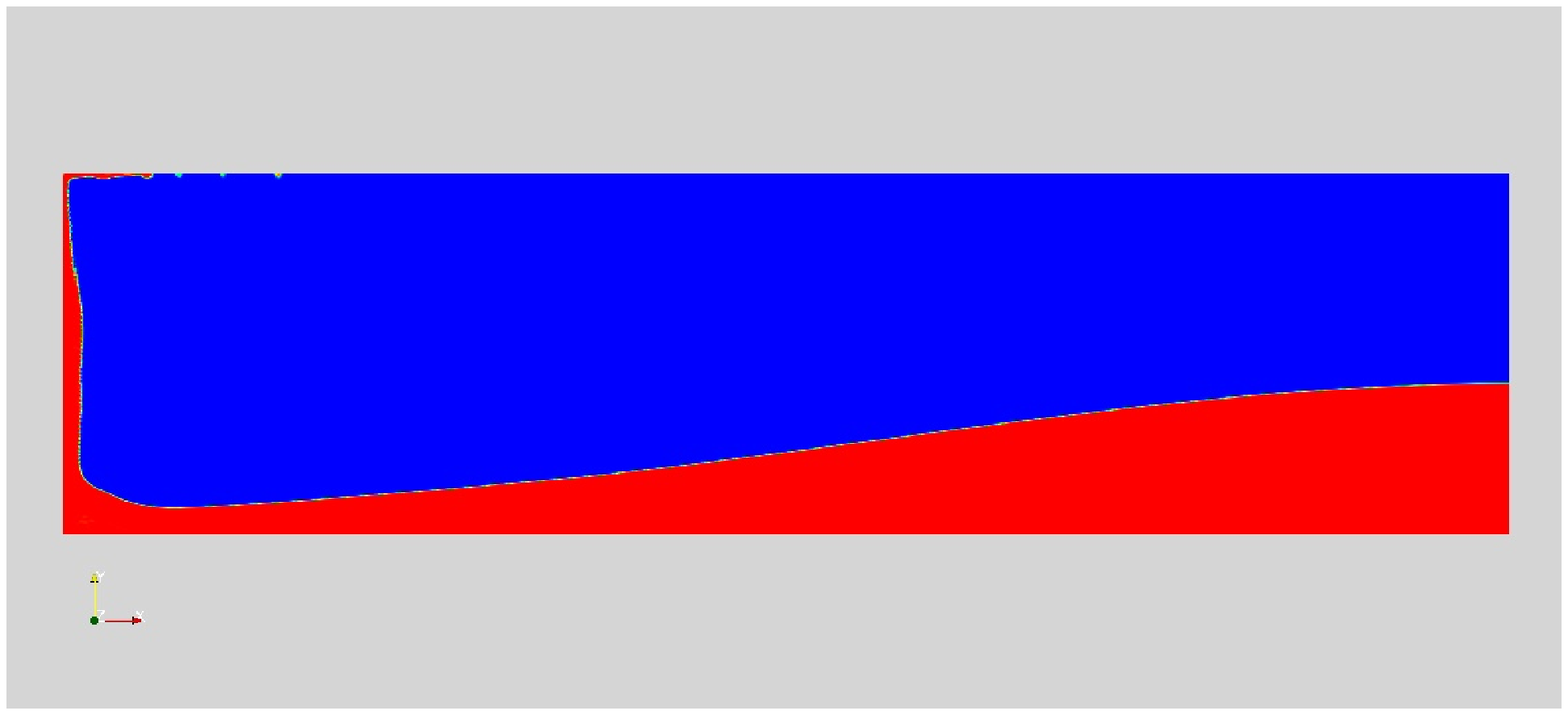}}
  \subfigure[\ac{NSWE} with \VOLNA solver]%
  {\includegraphics[width=0.39\textwidth]{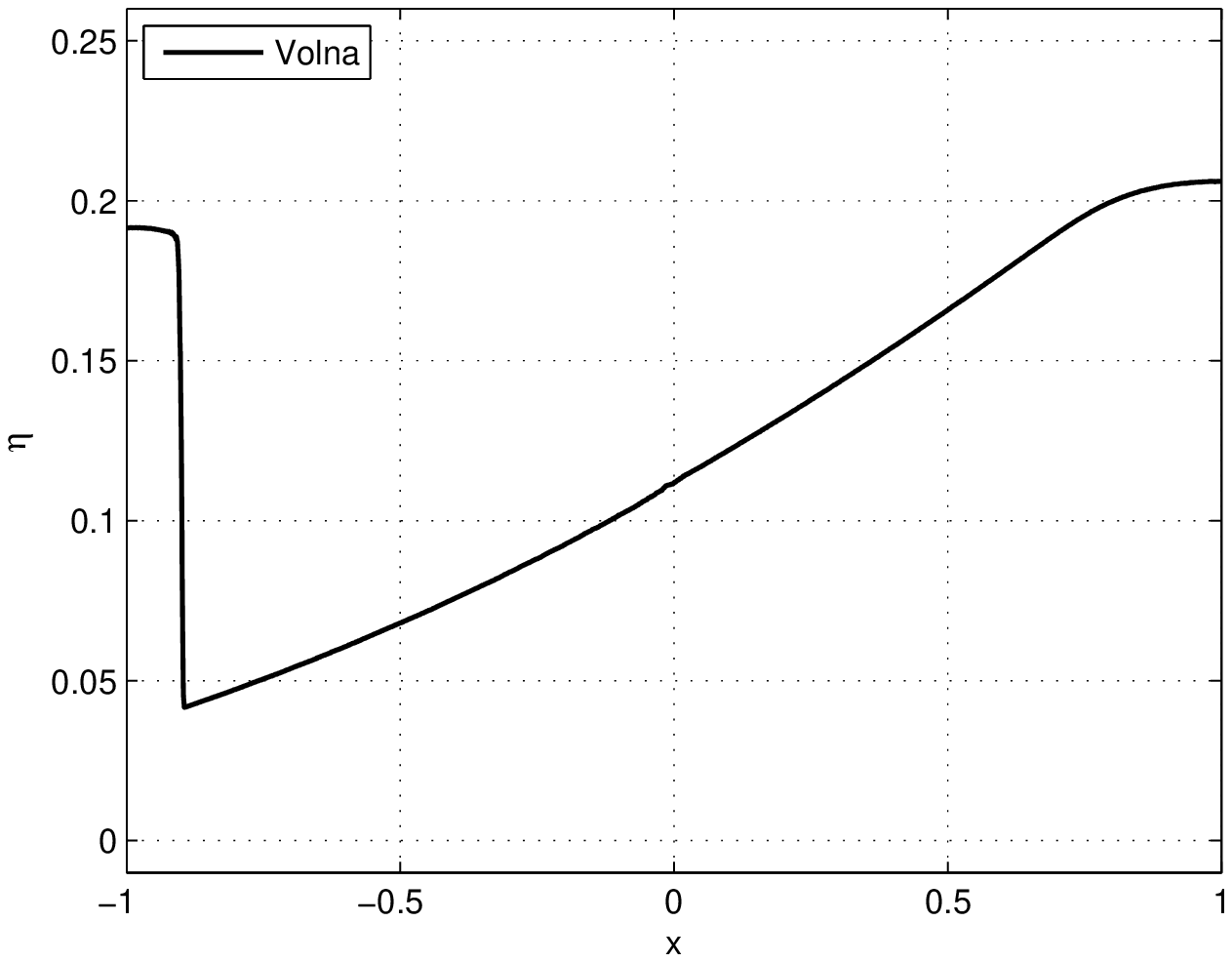}}
  \caption{Interaction with the left vertical wall at $t = 2.2$ s.}
  \label{fig:impact2}
\end{figure}

The comparison results are presented in Figures \ref{fig:impact1} and \ref{fig:impact2}. For instance, the wave amplitude on Figure \ref{fig:impact2} (a) reaches the upper boundary (its height is $0.5$ m), while \ac{NSWE} numerical solution amplitude does not exceed $0.2$ m.

From these results it is obvious that the wave impact process is not correctly modelled by \ac{NSWE}. It is possible to foresee this conclusion if one recalls two main constutive assumptions behind \ac{NSWE}:
\begin{itemize}
  \item The pressure is hydrostatic
  \item Vertical velocity and acceleration are neglected
\end{itemize}
Notice, that for infinitely long waves, these two hypotheses are equivalent. Since, the dynamic pressure dominates in the impact process, we get qualitatively wrong results (it is especially clear from Figure~\ref{fig:impact2}).

%%%%%%%%%%%%%%%%%%%%%%%%%%%%%%%%%%%%%%%%%%%%%%%%%%%%%%%%%%%

\section{Conclusions and perspectives}\label{sec:concl}

In the present study we tried to examine the validity of the \ac{NSWE} for wetting (flooding) process modelling. As a test-case we chose the classical dam break problem which is the \textit{de facto} standard in this field. This problem was solved in the context of two completely different models in terms of physical precision and, by consequence, of different complexity. The two-fluid \ac{DNS} was chosen as the reference solution since all necessary physical effects are included in it.

Comparison results presented above show good overall performance of the \ac{NSWE}. In order to appreciate more these results, one should take into account also the computational cost of the \ac{DNS} and relatively inexpensive shallow water simulations.

However, we revealed several drawbacks of the depth-integrated model. Namely, the free surface shape differs from the parabolic profile predicted by \ac{NSWE}. This discrepancy is attributed to the non-piecewise linear distribution of the velocity field inside the water column. To compare, see Figure \ref{fig:u_mag1_4} for the \ac{DNS} result and formula (\ref{eq:StokerU}) for the analytical prediction by \ac{NSWE}. The experimental and theoretical study of P.K. Stansby {\sl et al} (1998) \cite{Stansby1998} also revealed some differences during the initial stages. However, their objection concerned essentially some new jet-like phenomena just after release. For later times, they found relatively close agreement with \ac{NSWE}.

We went beyond the initial purpose of this study and continued our simulations until the impact process with the left wall. It was shown that the \ac{NSWE} strongly underestimate the wave height. This discrepancy has its origins in the hydrostatic pressure assumption. Actually, the dynamic pressure becomes dominant during the impact process. Its excess is responsible of spectacular splashes that we may have a chance to observe in nature.

Concerning the front propagation speed, we obtained slightly different values between the \ac{DNS} and the \ac{NSWE} solutions. We have to note that in a physical experiment this quantity strognly depends on the soil conditions. The standard no-slip boundary condition is clearly insufficient to describe all kinds of soils. We believe that future research activities will focus on developing wall function laws and realistic boundary conditions for Navier-Stokes equations (two-fluid or with free surface). On the other hand, \ac{NSWE} can also be improved. To produce physically correct results, these equations should be completed by friction laws (Ch\'ezy, Manning, Darcy-Weisbach and other laws) with properly adjusted coefficients. Recently, bottom boundary layer effects on long waves were studied \cite{Dutykh2008a}. Another direction consists in extending \ac{NSWE} to account for bed material transport as it was recently proposed by Fraccarollo \& Capart (2002) \cite{Fraccarollo2002}.

%%%%%%%%%%%%%%%%%%%%%%%%%%%%%%%%%%%%%%%%%%%%%%%%%%%%%%%%%%%

\section*{Acknowledgments}

The first author would like to acknowledge the support from ANR MathOc\'ean (Project n$^o$ ANR-08-BLAN-0301-01) and from the program ``Risques gravitaires, s\'eismes'' of Cluster Environnement and the research network VOR. The work of D.~Mitsotakis was supported by Marie Curie Fellowship No. PIEF-GA-2008-219399 of the European Commission. 

We would like to thank Fr\'ed\'eric Dias, Jean-Michel Ghidaglia and Rapha\"el Poncet from CMLA, ENS de Cachan for their invaluable participation in developing the operational code \VOLNA. Also, the authors would like to acknowledge Professors Costas Synolakis and Jean-Claude Saut for helpful discussions. Their works on water waves are limitless source of our inspiration. The first author is grateful to C\'eline Acary-Robert for her permanent assistance and valuable advice. Finally, special thanks go to David Lannes who organized the workshop ``Oceanography \& Mathematics'' (26 -- 28 January 2009 at ENS Paris) and brought together the authors.

%%%%%%%%%%%%%%%%%%%%%%%%%%%%%%%%%%%%%%%%%%%%%%%%%%%%%%%%%%%

%\bibliography{biblio}
%\bibliographystyle{alpha}
\newcommand{\etalchar}[1]{$^{#1}$}

\medskip
Received 21/03 2009; revised 24/05 2009.
\medskip

\end{document}